\documentclass[12pt]{article}
\usepackage{epsfig,cite}
\usepackage {amsmath}
\usepackage{amssymb}
\usepackage{color}
\usepackage{graphicx}
\usepackage{verbatim,mathrsfs,subfig,feynmf}
\include{epsf}
\newcount\timecount
\newcount\hours \newcount\minutes  \newcount\temp \newcount\pmhours
\hours = \time
\divide\hours by 60
\temp = \hours
\multiply\temp by 60
\minutes = \time
\advance\minutes by -\temp
\def\hour{\the\hours}
\def\minute{\ifnum\minutes<10 0\the\minutes
            \else\the\minutes\fi}
\def\clock{
\ifnum\hours=0 12:\minute\ AM
\else\ifnum\hours<12 \hour:\minute\ AM
      \else\ifnum\hours=12 12:\minute\ PM
            \else\ifnum\hours>12
                 \pmhours=\hours
                 \advance\pmhours by -12
                 \the\pmhours:\minute\ PM
                 \fi
            \fi
      \fi
\fi
}

\def\monthname{\relax\ifcase\month 0/\or January\or February\or
   March\or April\or May\or June\or July\or August\or September\or
   October\or November\or December\else\number\month/\fi}

\def\bold#1{\setbox0=\hbox{$#1$}%
     \kern-.025em\copy0\kern-\wd0
     \kern.05em\copy0\kern-\wd0
     \kern-.025em\raise.0433em\box0 }

\def\beq{\begin{equation}}
\def\eeq{\end{equation}}

\def\ga{\mathrel{\raise.3ex\hbox{$>$\kern-.75em\lower1ex\hbox{$\sim$}}}}
\def\la{\mathrel{\raise.3ex\hbox{$<$\kern-.75em\lower1ex\hbox{$\sim$}}}}
\def\gev{{\rm \, Ge\kern-0.125em V}}
\def\tev{{\rm \, Te\kern-0.125em V}}
\def\gyr{{\rm \, G\kern-0.125em yr}}

\def\gappeq{\mathrel{\rlap {\raise.5ex\hbox{$>$}}
{\lower.5ex\hbox{$\sim$}}}}
\def\lappeq{\mathrel{\rlap{\raise.5ex\hbox{$<$}}
{\lower.5ex\hbox{$\sim$}}}}
\def\Toprel#1\over#2{\mathrel{\mathop{#2}\limits^{#1}}}

\def\m12{m_{1\!/2}}

\def\bea{\begin{eqnarray}}
\def\eea{\end{eqnarray}}

\def\beq{\begin{equation}}
\def\eeq{\end{equation}}

\begin{document}

\begin{titlepage}
\pagestyle{empty}
\baselineskip=21pt
\rightline{UMN--TH--3314/13, FTPI--MINN--13/41, IPMU13-0227}
\vspace{0.2cm}
\begin{center}
{\large {\bf One-Loop Anomaly Mediated Scalar Masses \\and $(g-2)_\mu$ in Pure Gravity Mediation }}
\end{center}
\vspace{0.5cm}
\begin{center}
{\bf Jason L. Evans}$^{1}$,
{\bf Masahiro Ibe}$^{2,3}$ {\bf Keith~A.~Olive}$^{1}$
and {\bf Tsutomu T. Yanagida}$^{3}$\\
\vskip 0.2in
{\small {\it
$^1${William I. Fine Theoretical Physics Institute, School of Physics and Astronomy},\\
{University of Minnesota, Minneapolis, MN 55455,\,USA}\\
$^2${ ICRR, University of Tokyo, Kashiwa 277-8582, Japan}\\
$^3${Kavli IPMU (WPI), TODIAS, University of Tokyo, Kashiwa 277-8583, Japan}\\
}}
\vspace{1cm}
{\bf Abstract}
\end{center}
\baselineskip=18pt \noindent
{\small We consider the effects of non-universalities among sfermion generations in models
of Pure Gravity Mediation (PGM). In PGM models and in many models with
strongly stabilized moduli, the gravitino mass may be O(100) TeV, whereas gaugino
masses, generated through anomalies at 1-loop, remain relatively light O(1) TeV.
In models with scalar mass universality, input scalar masses are generally very heavy
($m_0 \simeq m_{3/2}$) resulting in a mass spectrum resembling that in split
supersymmetry. However, if one adopts a no-scale or partial no-scale structure
for the K\"ahler manifold, sfermion masses may vanish at the tree level. It is usually
assumed that the leading order anomaly mediated contribution to scalar masses
appears at 2-loops. However, there are at least two possible sources for 1-loop
scalar masses. These may arise if Pauli-Villars fields are introduced as messengers  of
supersymmetry breaking.  We consider the consequences of a spectrum in which
the scalar masses associated with the third generation are heavy (order $m_{3/2}$) with
1-loop scalar masses for the first two generations. A similar spectrum is expected to
arise in GUT models based on $E_7/SO(10)$ where the first two generations of scalars
act as pseudo-Nambu-Goldstone bosons. Explicit breaking of this symmetry by the gauge couplings then generates one-loop masses for the first two generations.
In particular, we show that it may be possible
to reconcile the $g_\mu - 2$ discrepancy with potentially observable
scalars and gauginos at the LHC.}


\vfill
\end{titlepage}

\section{Introduction}
Although the mass of the recently discovered Higgs boson\cite{lhch} is light enough that it can be accommodated in supersymmetry, it is near the upper limit of simple models like the CMSSM\cite{cmssm,cmssmh}. This large Higgs mass and the lack of evidence for supersymmetric particles at the LHC \cite{lhc,ATLASsusy} have put severe constraints on the simplest models of supersymmetry \cite{125-other,pressure} including the CMSSM.

Since both the LHC constraints on the superpartners and the observed Higgs mass favor heavier sfermion masses \cite{lhc,ATLASsusy}, it could be that nature does indeed have a mass splitting among the supersymmetric particles as is the case in split supersymmetry\cite{split}, pure gravity mediation (PGM) \cite{pgm,pgm2,eioy,eioy2}, and models with strongly stabilized moduli\cite{klor,Dudas:2006gr,dlmmo}. In models of PGM\cite{pgm}, sfermions get a tree-level mass, as in the CMSSM, while gauginos get a one-loop mass from anomaly mediation\cite{anom}.  Recently, we showed that models based on Pure Gravity Mediation, with \cite{eioy} and without \cite{eioy2} scalar mass universality, could explain virtually all experimental constraints with electroweak symmetry breaking generated radiatively.  In the case of full scalar mass universality, the theory can be described in terms of two free parameters, the gravitino mass, $m_{3/2}$ and $\tan\beta$ the ratio of the Higgs vacuum expectation values.  However, these models placed a rather strict constraint on $\tan\beta=1.7-2.5$. The Higgs mass constraint then restricted the gravitino mass to the range $m_{3/2}=300-1500$ TeV. If the Higgs soft masses are allowed to be non-universal, $\tan\beta$ is only restricted by perturbativity of the Yukawa couplings and $m_{3/2}$ can be as low as $80$ TeV. However, even for a gravitino mass this light all sfermions masses are much larger than the weak scale.

If all sleptons have mass of order $80$ TeV or more, there is little hope of explaining the discrepancy in the anomalous magnetic moment of the muon \cite{newBNL} or sfermion detection at the LHC. As was recently shown in \cite{bingm2}, sleptons need to be lighter than about $2$ TeV if there is to be any hope of explaining $(g-2)_\mu$.  The LHC reach varies greatly depending on the masses of the first two generation squark masses. If squarks are lighter than $2$ TeV, the LHC reach on the gluino can be as high as about 4 TeV\cite{AtlasNote}.  To get sfermion masses this light in PGM, there must be additional sources of non-universalities in the sfermion boundary masses. Since large stop masses are important in explaining the Higgs mass \cite{mh}, it will be advantageous to take tree-level masses of order $m_{3/2}$ for the stops.   Furthermore, if the Higgs bi-linear mass term, $\mu$, is much larger than the stau mass, as is often the case in PGM, the stau tends to be tachyonic \cite{bingm2}. This problem can also be evaded by having a tree-level stau mass.  These arguments suggests that the third generation should have tree-level masses while the first and second generations boundary masses should be suppressed.  Phenomenologically viable models can also be found for suppressed third generation masses, however, they tend to be qualitatively similar to the PGM models discussed in \cite{eioy,eioy2}.

A, possibly, more compelling reason to discuss light first and second generation sfermion masses is the hierarchy in the Yukawa couplings. If the first two generations are pseudo Nambu-Goldstone multiplets (pNG) of some broken global symmetry \cite{Buchmuller:1983na}, this would naturally suppress the sfermion masses.  Since the Yukawa couplings are an explicit breaking of the global symmetry, the Yukawa couplings of the pNG would also be suppressed.  A similar suppression of the first and second generation sfermion masses can be realize from a no-scale like geometry for the K\"ahler potential \cite{Cremmer:1983bf}.  This geometry can arise from a brane separation where on one brane we have the SUSY breaking fields as well as the Higgs boson and third generation fields and on the other we have the first and second generations fields. In both of these scenarios the Yukawa coupling hierarchies are linked to the sfermion mass hierarchies.

Generating hierarchically small soft masses for the first two generations is not so problematic. However, because the gauginos are small in comparison to the third generation masses, the RG running of the first two generations will give tachyonic masses for the simplest of models.  These tachyonic masses can be evaded if sfermion masses of the first two generations are generated at one-loop.  In the case of no-scale like boundary conditions this can be accomplished if the Pauli-Villars fields, that regularize the low-scale theory, interact with supersymmetry breaking generating a one-loop soft masses\cite{oneloopmass}. The Pauli-Villars fields act as the messengers of supersymmetry breaking. In the case of $E_7/SO(10)$\cite{eioy2}, the preons act as messengers generating a similar one-loop mass much like the Pauli-Villar fields. Thus, it is possible that we can construct a spectrum in which $m_{{\tilde u},{\tilde c}} \sim m_{\tilde g} \ll m_{\tilde t} \sim m_{3/2}$, where  $m_{{\tilde u},{\tilde c},{\tilde t}}$ refer to the three generations of sfermion masses, and $m_{\tilde g}$ refers to gaugino masses. As we will see, this type of mass hierarchy is capable of simultaneously explaining the Higgs mass and the deviation in $(g-2)_\mu$.

In section 2, we will discuss our model of PGM which will allow for light first and second generation sfermions. We also describe the mechanism for generating one-loop anomaly mediated masses for the first two generation sfermions.  As we will see, due to our ignorance of the precise mechanism for
transmitting supersymmetry breaking, we inevitably have three new parameters associated with the
one-loop masses correlated with the three low energy gauge groups. In Section 3, we derive results
with light first and second generation sfermions in the context standard grand unified theories in which there is an assumed relation between the new parameters, and in section 4 we discuss the impact of these models on the value of deviation in the anomalous magnetic moment of the muon,
$\Delta {\rm a}_\mu$.  In section 5,
we will discuss alternate grand unified scenarios where the anomalous magnetic moment of the muon can be more easily explained. Lastly, in section 6 we will conclude.

\section{PGM and More Non-Universalities}
The back bone of our discussion will be the pure gravity mediated models discussed in \cite{eioy,eioy2} with a K\"ahler potential
\begin{eqnarray}
K =   y_i y_i^*  + K^{(H)} +  K^{(Z)}  +  \ln |W|^2\ ,
\end{eqnarray}
where the K\"ahler potential for the Polonyi-like modulus, $Z$, which is responsible for supersymmetry
breaking, contains a stabilizing term \cite{dine}
\begin{eqnarray}
K^{(Z)} = Z Z^* \left(1-\frac{Z Z^*}{\Lambda^2}\right) \ ,
\end{eqnarray}
and the K\"ahler term for the Higgs fields contains a Giudice-Masiero-like term \cite{gm,dmmo,eioy}
\begin{eqnarray}
K^{(H)}=|H_1|^2+|H_2|^2+c_H\left(H_1H_2+c.c\right) \, ,
\label{kh1}
\end{eqnarray}
and $y_i$ represent the other MSSM fields. We also assume that the superpotential is separable
between the matter fields and hidden sector fields:
\beq
W = W^{(Z)} + W^{({\rm SM})} \, ,
\eeq
where $W^{({\rm SM})}$ contains all Standard Model (SM) contributions to the superpotential.
Furthermore, we assume a simple Polonyi form for $W^{(Z)}$ \cite{pol},
\beq
W^{(Z)}= \tilde m^2(Z+\nu),
\eeq
It has recently been shown that strongly stabilized models of this type
are free from any of the cosmological problems normally associated with
moduli or gravitinos if $\Lambda \lesssim 3 \times 10^{-4}$  \cite{ego}.

 For this K\"ahler potential, the MSSM scalar fields will have an input mass $m_{\tilde f}=m_{3/2}$
 at the universality scale which we associate with the Grand Unified Theory (GUT) scale. In the absence of a non-trivial gauge kinetic function, the gaugino masses are generated from anomalies and will have loop suppressed masses given by
\begin{eqnarray}
M_i=b_i\frac{g_i^2}{16\pi^2} m_{3/2} \, ,
\end{eqnarray}
where the $b_i$ are the coefficients of the beta function.  The tree-level contribution to the $A$-terms are quite small, $A_0\sim (\Lambda^2/M_P^2)m_{3/2}$ \cite{dlmmo}. The leading order contribution to the $A$-terms are the one-loop anomaly mediated contributions and are effectively zero.

For the universal case discussed above, $\tan\beta$ is restricted to the range $1.7-2.5$ which forces $m_{3/2}\gtrsim 300$ TeV in order to get a sufficiently large Higgs mass \cite{eioy}.
However, if we take non-universal Higgs boundary masses \cite{eioy2}, $\tan\beta$ is only constrained by the weaker restrictions of perturbativity of the Yukawa couplings.
Non-universality is easily achieved by adding non-minimal couplings of the Higgs fields
to the modulus, $Z$. For example,
\begin{eqnarray}
K^{(H)}=  \left(1 + a \frac{Z Z^*}{M_P^2}\right) H_1 H_1^*
+ \left(1 + b \frac{Z Z^*}{M_P^2}\right) H_2 H_2^* + (c_H H_1 H_2 + h.c.)
\label{kh2}
\end{eqnarray}
will generate Higgs soft masses which depend on the couplings $a$ and $b$ \cite{eioy2}
\beq
m_1^2 = (1 - 3a) m_{3/2}^2; \qquad m_2^2 = (1 - 3b) m_{3/2}^2 .
\eeq
In this case, the lower bound on $m_{3/2}$ is due to the wino mass \cite{ATLASwino} placing a lower bound of about $m_{3/2}\simeq 80$ TeV.

The RG running in these models is rather simple.  Since the gaugino masses are small, they do not affect the RG running of the sfermion masses. Because only the third generation Yukawa couplings are large, only the third generation masses will run at one-loop.  However, the variations of the third generation masses from RG running preserves ${\cal O}(m_{3/2})$ masses for the third generation. If all the sfermion masses are ${\cal O}(m_{3/2})$, they cannot be seen at the LHC and will be of no aid in explaining the discrepancy in $(g-2)_\mu$. To make things worse, if all scalar masses are universal at the GUT scale, their masses need to be or order $300$ TeV to get a suitably large Higgs mass. Only by breaking the universality of the Higgs soft masses can this constraint on $\tan\beta$  be weakened. The lower bound on the scalar masses can then be as low as $80$ TeV, with this lower bound coming from the constraints on the wino mass. But, even sfermion masses of order $80$ TeV can not explain $(g-2)_\mu$ or be detected at the LHC.

To have anything other than the vanilla gauginos signals at accelerators for these models, at least some of the scalars need to be light and thus additional non-universalities are needed beyond the Higgs soft masses. As is well known \cite{anom}, in the absence of a large tree level scalar mass, scalar
masses are present at least at the two-loop level. However, as we discuss in more detail below,
it is possible that scalar masses also arise at one-loop.
Indeed one can imagine a no-scale construction where all scalar masses vanish at the tree level
as in no-scale supergravity \cite{Cremmer:1983bf}.  The K\"ahler potential can be written as
\begin{eqnarray}
K= -3 \ln\left(1-\frac{1}{3}\left[K^{(Z)}+K^{(H)}+y_iy_i^*\right]\right)+\ln |W|^2\ ,
\end{eqnarray}
where $K^{(H)}$ is given by Eq. (\ref{kh1}).  If all sfermion masses vanish at the tree level
and receive one-loop contributions, it will be difficult to generate a Higgs mass
as large as 125 GeV for generic parameters unless $m_{3/2}\gtrsim 150$ TeV. Since the sfermions are still rather heavy, this model will be qualitatively the same as PGM.

Instead, the approach we take below is to suppress only the masses of the first and second generation sfermion masses. Here, we discuss two ways of suppressing scalar masses of the first two generation sfermions.  The first  is to take a similar no-scale like K\"ahler potential of the form
\begin{eqnarray}
K=y_i^{(3)}y_i^{*(3)} -3 \ln\left(1-\frac{1}{3}\left[K^{(Z)}+y_i^{(1,2)}y_i^{*(1,2)}\right]\right)+K^{(H)}+\ln |W|^2\ ,
\end{eqnarray}
where $y_i^{(1,2)}$ are first and second generation fields in the MSSM and $y_i^{(3)}$ are the third generation fields. Although this K\"ahler potential is capable of suppressing the sfermion masses, it will be advantageous to also take non-universal Higgs masses coming from a K\"ahler potential of the form
given in Eq. (\ref{kh2}).
For this model, the bulk of the features of PGM remain but in addition we have very light sfermion masses for the first two generations which are now generated by anomalies.

The other possibility for suppressing the first and second generation sfermion masses is to associate these fields with the pNG of the global symmetry $E_7/SO(10)$. However, in this case the gauge and Yukawa couplings act as an explicit breaking of this symmetry. As we will see below, this is actually an advantage. The gauge and Yukawa couplings break the symmetry and one-loop masses are generated.

In an actual no-scale like model, the sfermion masses would be generated from the one-loop gaugino mass contributions to the RG equations. However, this no-scale like running is broken by the presence of a heavy third generation. This breaking of the no-scale structure has a drastic effect on the spectrum and as we will see,  we will need to find an additional source of mass for the first and second generation sfermions.

\subsection{General Features of the Renormalization Group Running}
In this section we will discuss the bulk features of the running of the first and second generation sfermion masses. As usual, we can take the third generation dominance approximation and will neglect the Yukawa couplings of the first two generations (see appendix \ref{sec:fcnc} on the SUSY FCNC contributions).  In this approximation, the only one-loop contribution to the first two generation sfermion mass running comes from gaugino masses and
\begin{eqnarray}
S=\frac{1}{2} {\bf Tr}\left(Ym^2\right) \, ,
\end{eqnarray}
where $Y$ is the hypercharge and $m^2$ represents the sfermion masses of the particles charged under hypercharge. Since this contains contributions from the third generation, it will generally be the dominant contribution to the running of the first two generations. The change in the sfermions masses from $S$ can be easily determined because it has a rather simple RG equation,
\begin{eqnarray}
\frac{d S_Y}{dt}= \frac{g_Y^2}{8\pi^2} \sum\limits_i \left(\frac{Y_i}{2}\right)^2S_Y \, ,
\end{eqnarray}
with solution,
\begin{eqnarray}
S_Y(Q) = S_Y(Q_0) \frac{g_Y^2(Q)}{g_Y^2(Q_0)} \, .
\end{eqnarray}
After integrating the RG, this contribution to the sfermion masses is of order ${\cal O}(m_{3/2})$.  This is much too large and would typically lead to tachyonic sfermion masses.  However, if sfermion masses are universal or determined by gauge interactions, $S_Y(Q_0)=0$ and so it remains zero at one-loop for the entire running\footnote{This relationships is broken at two-loops. However the effect of $S_Y$ still tends to be sub-dominant in this case.}. $S_Y(Q_0)=0$ is unchanged for non-universal Higgs masses as long as $m_1^2=m_2^2$, as in the NUHM1 \cite{nuhm1}.  Since we are considering a combination of these models, we have $S_Y(Q_0)=0$ and $S_Y$ does not play a significant role in the RG running, though it is included in our analysis below.

As stated above, the other one-loop contribution to the RG running of the first two generation is proportional to the gaugino masses squared. Since the gaugino masses are loop suppressed relative to $m_{3/2}$, their effective contribution to the RG running is of order
\begin{eqnarray}
\frac{m_{3/2}^2}{(16\pi^2)^3},
\end{eqnarray}
effectively a three-loop contribution much too small to be important. Thus, the two-loop contributions which are proportional to third generation masses will have a much more important effect on the masses of the first and second generation sfermions.

Since the tree-level sfermion masses of the first two generations are suppressed, terms in the beta functions proportional to them will not be important.  Only contributions involving third generation masses are significant.  Generation mixing in the RG running is through loops of $D$-terms, i.e. RG terms coming from $\langle (D^aD^a)^2 \rangle $ or $\langle D^aD^a (\tilde f_i f_jf_k)^2\rangle $ which give terms like those in Eq. (\ref{eq:twoloopY})--(\ref{eq:twoloopSp}). The rough sizes of these contributions to the RG running of the first two generations are
\begin{eqnarray}
{\cal O}(1) \frac{g_i^4}{(16\pi^2)^2}m_{3/2}^2 \quad\quad {\rm and }\quad\quad  {\cal O}(1) \frac{g_1^2y_i^2}{(16\pi^2)^2}m_{3/2}^2 \, ,
\end{eqnarray}
where $g_i$ are the gauge couplings and $y_i$ are the Yukawa couplings. Their exact form can be found in Appendix \ref{sec:RGApp}.  As can be seen there, the RG running from a two-loop contribution in the beta function will diminish the sfermion mass by an amount of order
\begin{eqnarray}
{\cal O}(1) \frac{m_{3/2}^2}{(16\pi^2)},
\end{eqnarray}
if we are running down from the GUT scale.  Clearly, a one-loop boundary mass is need to offset the RG contribution to the mass and the two-loop anomaly mediated contribution is insufficient.

\subsection{Generating One-Loop Sfermion Masses}
In this section, we address the generation of one-loop masses for the sfermions. Since string theory is a renormalizable theory, it should provide some mechanism to renormalize itself.  The renormalization for the gauge interactions can be parameterized by adjoint Pauli-Villars (PV) fields.  Because string theory gives us no indication of how these PV fields interact with the hidden sector, we cannot say how strongly they feel supersymmetry breaking.  If the PV fields do in fact interact with the hidden sector they would act as messengers of supersymmetry breaking.  As was shown in \cite{oneloopmass}, this gives a one-loop contribution which is proportional to the gauge interactions and Yukawa couplings. In Appendix \ref{sec:toy}, we give a toy model showing how these one-loop masses are generated in the flat supersymmetric limit.  Since there is no way of knowing how the PV fields interact with the hidden sector, the masses of the sfermions are effectively free parameters.  However, we make the assumption that the PV fields corresponding to each generation interact with the hidden sector identically.  We find this a reasonable assumption since gauge symmetries in general do not distinguish between generations.

Another possibility is to consider a global $E_7/SO(10)$ which has two generations that are pNG. To have exact Nambu-Goldstone bosons (NGB), the gauge and Yukawa couplings need to be zero. By introducing explicit breaking to the $E_7/SO(10)$ in the form of gauge and Yukawa couplings, the masses of the NGB are lifted. These mass corrections should be at the one-loop order. This can be understood by noting that when the gauge interactions are turned on, they will generate one-loop corrections to the K\"ahler potential.  This one-loop correction deforms the K\"ahler potential of $E_7/S(10)$ breaking the cancellation needed to give massless fields. Since this breaking is at the one-loop order, we expect the sfermion masses to be generated at the one-loop order. As before, we get one-loop masses for the first and second generation sfermions. To calculate these  masses exactly we need the details of the underlying QCD like theory at the preon level. However, we know they are at the one-loop level and proportional to the gauge and Yukawa couplings.

\subsection{Parameterization}

To parameterize our lack of knowledge about the Plank scale dynamics or preon model, we will define

\begin{eqnarray}
\gamma_i = \frac{1}{8\pi^2}g_i^2 C(r) \, ,
\end{eqnarray}
where $g_i$ is the gauge coupling and $C(r)$ is the quadratic Casimir\footnote{The Casimir is important because we have adjoint fields interacting with fundamental fields in the superpotential. This will lead to a Casimir when we form loops from these interactions as can be seen in \cite{oneloopmass}.}. The soft mass for a given sfermion is then given by
\begin{eqnarray}
m_{\tilde f}^2= \sum\limits_i c_i \gamma_i m_{3/2}^2,
\end{eqnarray}
where we have made the assumption that the $c_i$ are generation independent.  Including these parameters, our full list of free parameters is
\begin{eqnarray}
m_{3/2}\quad \tan\beta \quad  m_1 = m_2 \quad c_1 \quad c_2 \quad c_2 .
\end{eqnarray}
The boundary masses for the first two generations then take the form
\begin{eqnarray}
m_{\tilde f _i}^2(Q_{GUT})= \sum\limits_jC_j(r_i)c_j\frac{g_j^2}{8\pi^2} m_{3/2}^2 \, ,
\end{eqnarray}
where $c_j$ is defined above and $C_j(r_i)$ is the quadratic Casimir for $\tilde f_i$ from the gauge group $j$.

\section{Simple Unification}
We are now in a position to examine the simplest realization of this model, namely with $c_1=c_2=c_3$. This relationship among the $c_i$ is what would be expected if the grand unified theory stemmed from a simple $SU(5)$.  In this case the PV fields will stem from complete multiplets of the $SU(5)$ gauge group. If $SU(5)$ is broken in a generic fashion, we get $c_1=c_2=c_3$. In models like these, the lightest sfermion is a squark. Because the gauge couplings are universal at the GUT scale, where we apply our boundary masses, the squarks are only slightly heavier then the sleptons.  However, the RG running of the squarks is much stronger since $g_3\gg g_1$ at the weak scale. This leads to the lightest sfermion being the down squark as we explain below.

It is also important to note that non-universal Higgs soft masses are advantageous.  If $m_{1,2}^2\sim m_{3/2}^2$, we have $m_{3/2}\gtrsim 300$ TeV \cite{eioy} and even one-loop sfermion masses will remain out of reach for the LHC since generically $m_{\tilde q}$ would still be rather heavy. Not only would taking non-universal Higgs masses allow us to choose smaller $m_{3/2}$, it also has an important effect on the running. The non-universalities in the Higgs masses become important, because ${\cal S}'$  (listed in Appendix \ref{sec:RGApp} ) depends on the Higgs soft masses.  If the Higgs soft masses are universal, ${\cal S}'$ is suppressed and it has little affect on the running of the sfermion masses.  Because universality is not an option, we have a significant contribution to the sfermion mass running from ${\cal S}'$. This running splits the squark masses.

\begin{figure}[h!]
\subfloat[ ]{
{\includegraphics[width=0.5\columnwidth]{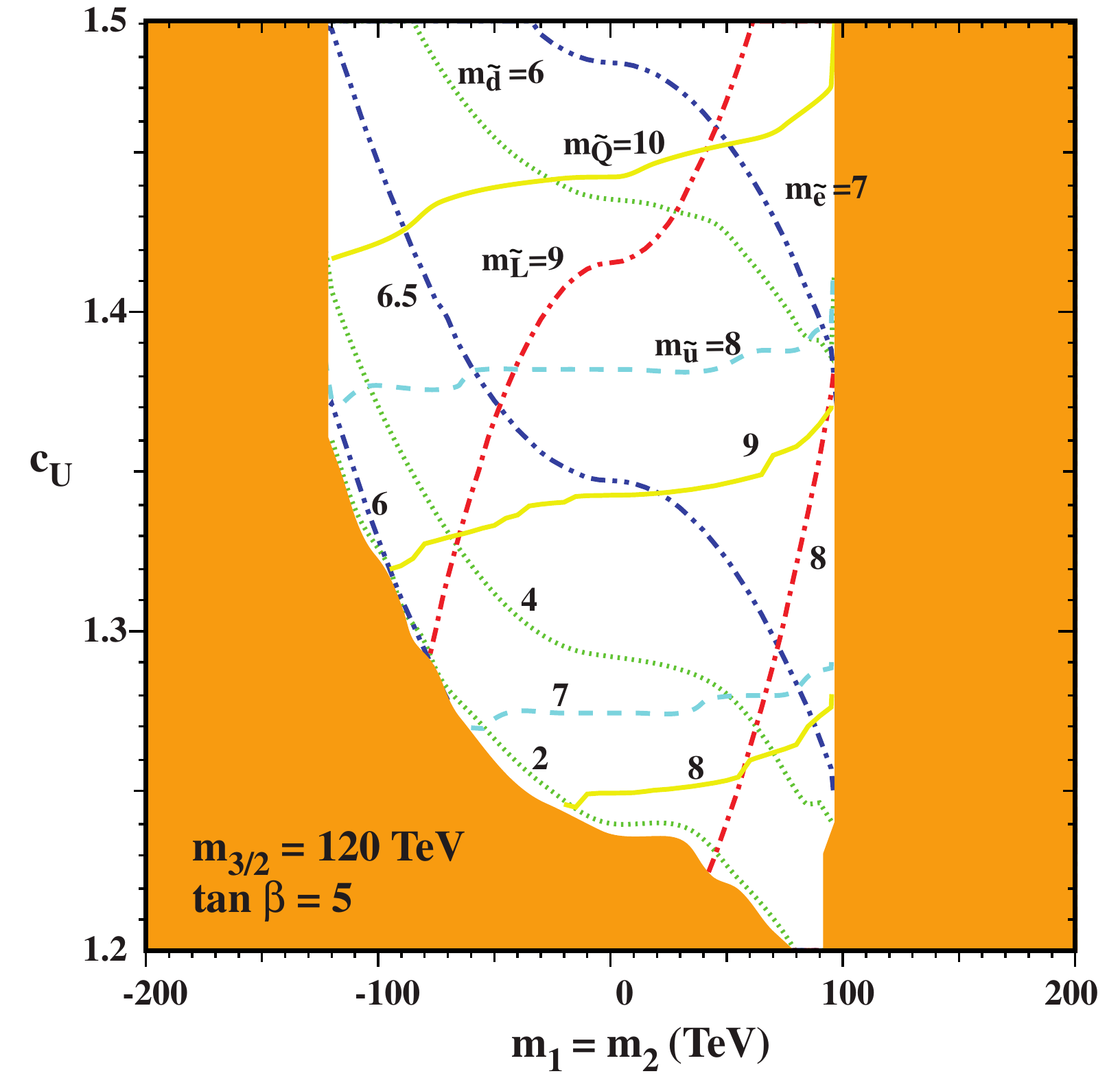}} }
\subfloat[]{
{\includegraphics[width=0.5\columnwidth]{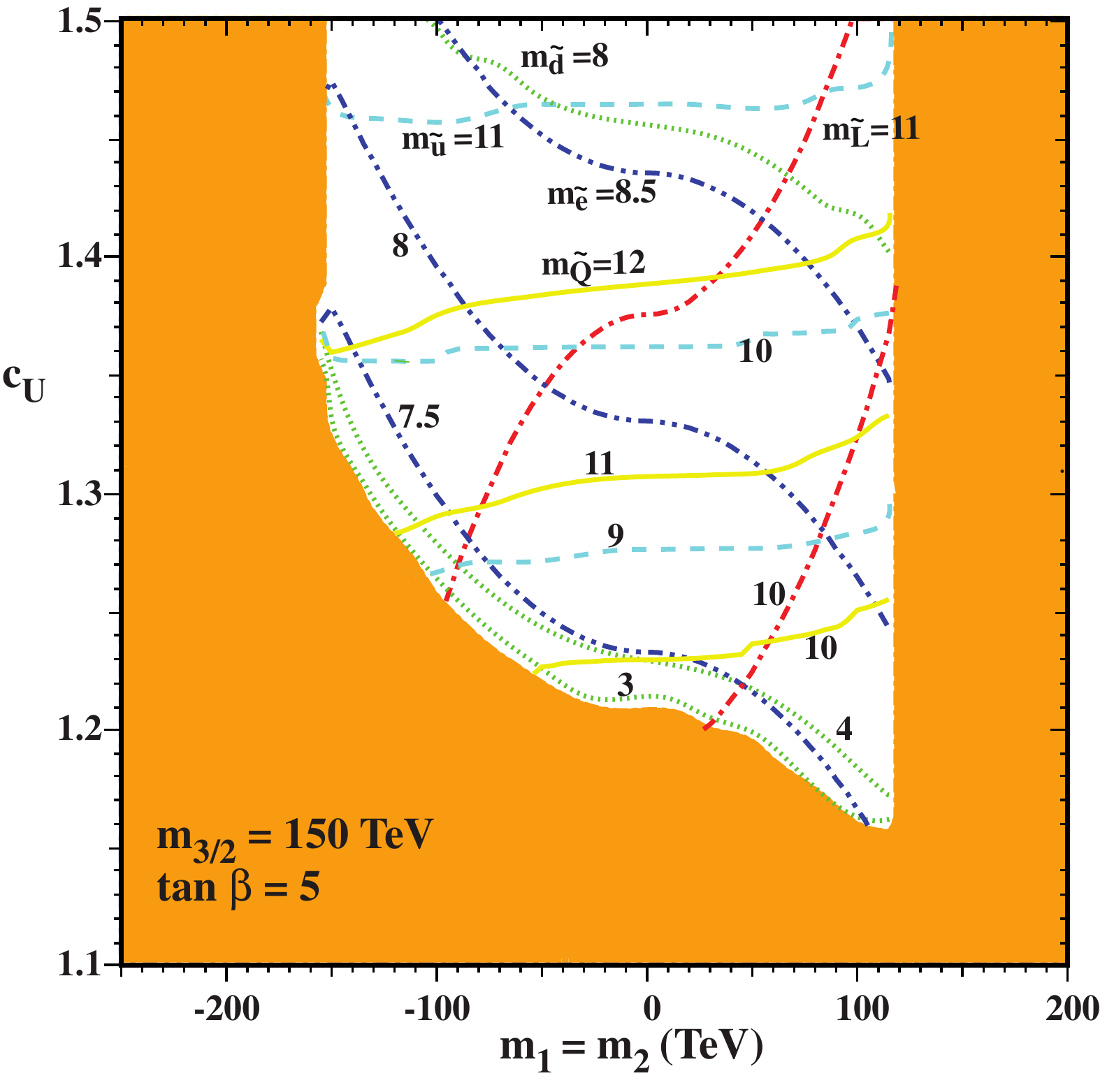}}}
\caption{ \it Here we show the contour plots of the sfermion masses of the first and second generation masses in the $m_1=m_2$ versus $c_U$ plane for (left) $m_{3/2} =120$ TeV and (right) $m_{3/2}=150$ TeV and $\tan\beta=5$. The line types are as follows: left-handed squarks (yellow solid);  right-handed scalar up (light blue dashed); right-handed scalar down (light green dotted); left-handed sleptons (red dot-dashed); and right-handed sleptons (blue double-dot-dashed) . The masses are in TeV. \label{fig:mhcu} }
\end{figure}

For the simplified model we consider here, the down squark is the lightest sfermion because it has the largest positive hypercharge.  With non-universal Higgs masses, ${\cal S}'$ is large and deflects the mass of $Q,d$ down and $u$ up. Since the hypercharge of $d$ is larger than that of $Q$, the down squark is the lightest. A plot of the mass spectra for these models can be seen in Fig. (\ref{fig:mhcu}) which shows the sfermion mass contours in the $m_1=m_2, c_U$ plane, where $c_U = c_1 = c_2 = c_3$ is the universal coefficient of the one-loop input
soft masses. The line type identifications are given in the caption. The shaded regions correspond to theoretically excluded regions for the following reasons: the upper left corner is exclude because $m_A^2 <0$, the lower region is excluded because scalar down is tachyonic. Notice that the down squark mass gets small near this boundary and the mass squared evolves very quickly as the boundary is approached, rapidly turning negative. The  shaded region on the right is excluded because $\mu^2 <0$. As can be seen in these figures, the down squark (green dashed curves) is the lightest sfermion.

To see the effect of changing $m_{3/2}$ we display, in Fig. (\ref{fig:mhcu}), two values of $m_{3/2}$, $m_{3/2}=120$ TeV (left) and $m_{3/2}=150$ TeV (right) for $\tan\beta=5$, the latter is chosen to get an acceptable Higgs mass. The sign for $m_1=m_2$ refers to the sign of $m^2$. As can be seen from the figures, the region with small down squark masses is shrinking and so it becomes increasingly more difficult to get a small mass for the down squark as $m_{3/2}$ becomes larger.  Once $m_{3/2}\gtrsim 100$ TeV, some degree of fine tuning is need to get sfermion masses less than about 2 TeV. The reason for this can be understood by examining the beta function.  As was discussed earlier, the leading order contribution to the beta function arises at two-loops and is proportional to gauge couplings. Since the third generation masses run very little, the beta function for the first two generations remains fairly constant and are of order
\begin{eqnarray}
\beta_{m_{\tilde f}^2} \sim {\cal O}(1) \frac{g_i^2}{(16\pi^2)^2}m_{3/2}^2\sim \left(2~ {\bf TeV}\right)^2 \left(\frac{m_{3/2}}{100 ~{\bf TeV}}\right)^2 \, .
\end{eqnarray}
Once the sfermion masses become similar in size to the beta function, the sfermion mass will be quickly driven to zero\footnote{In this regime, the typical approximation of setting $m_{\tilde f}=0$ in the beta functions once $m_{\tilde f}< Q$, where $Q$ is the RG scale, is invalid. In the figures, we assume that we can extrapolate between regions where we can safely integrate out the sfermions to the region where the sfermions become  tachyonic, knowing that all possible sfermion masses should be traversed. } .  Thus, even if we adjust the boundary mass of the sfermions, it will be difficult to get a sfermion mass smaller than the size of the beta function.

\begin{figure}[t!]
\begin{center}
\subfloat[]{\includegraphics[clip, width=0.5\columnwidth]{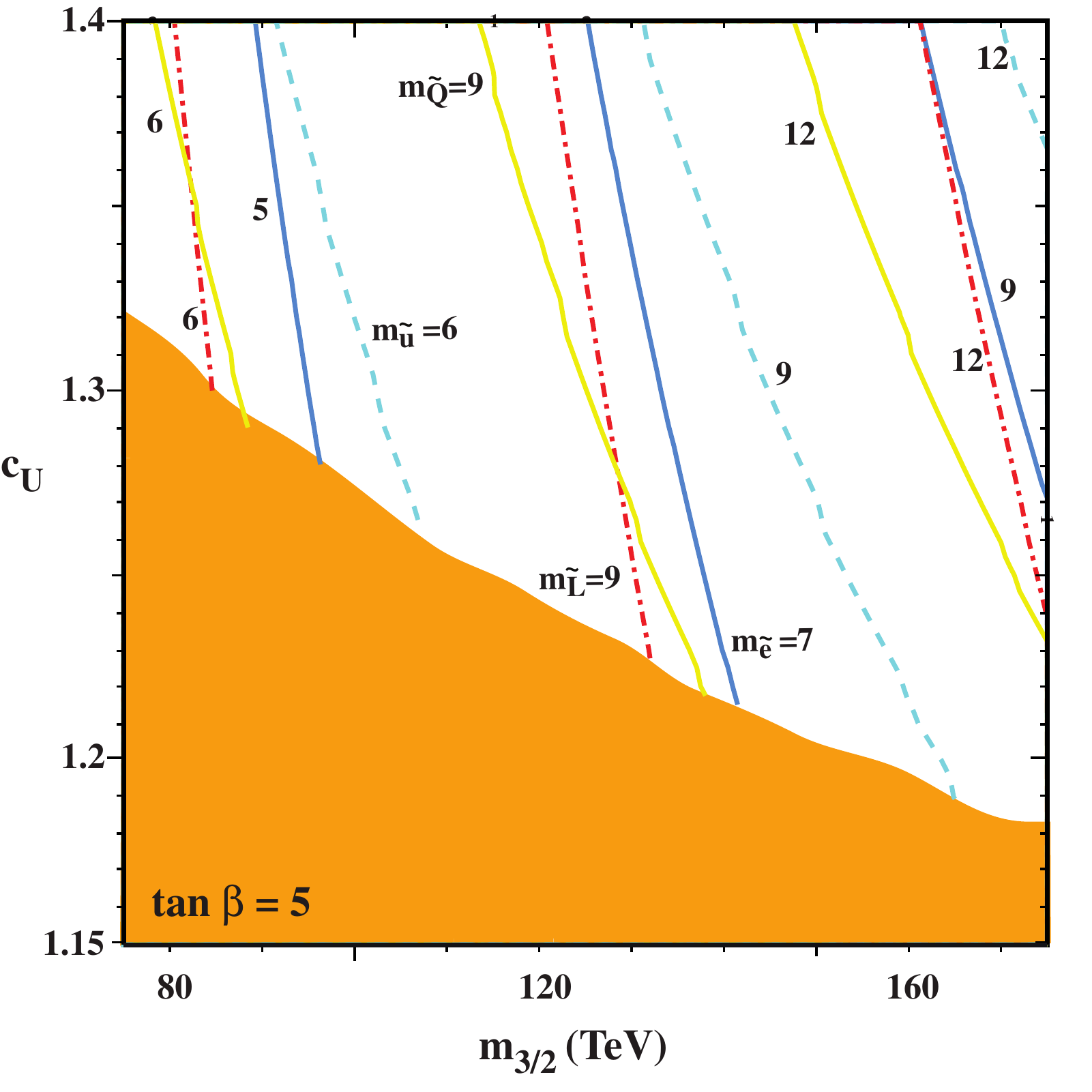}}
\subfloat[]{\includegraphics[clip, width=0.5\columnwidth]{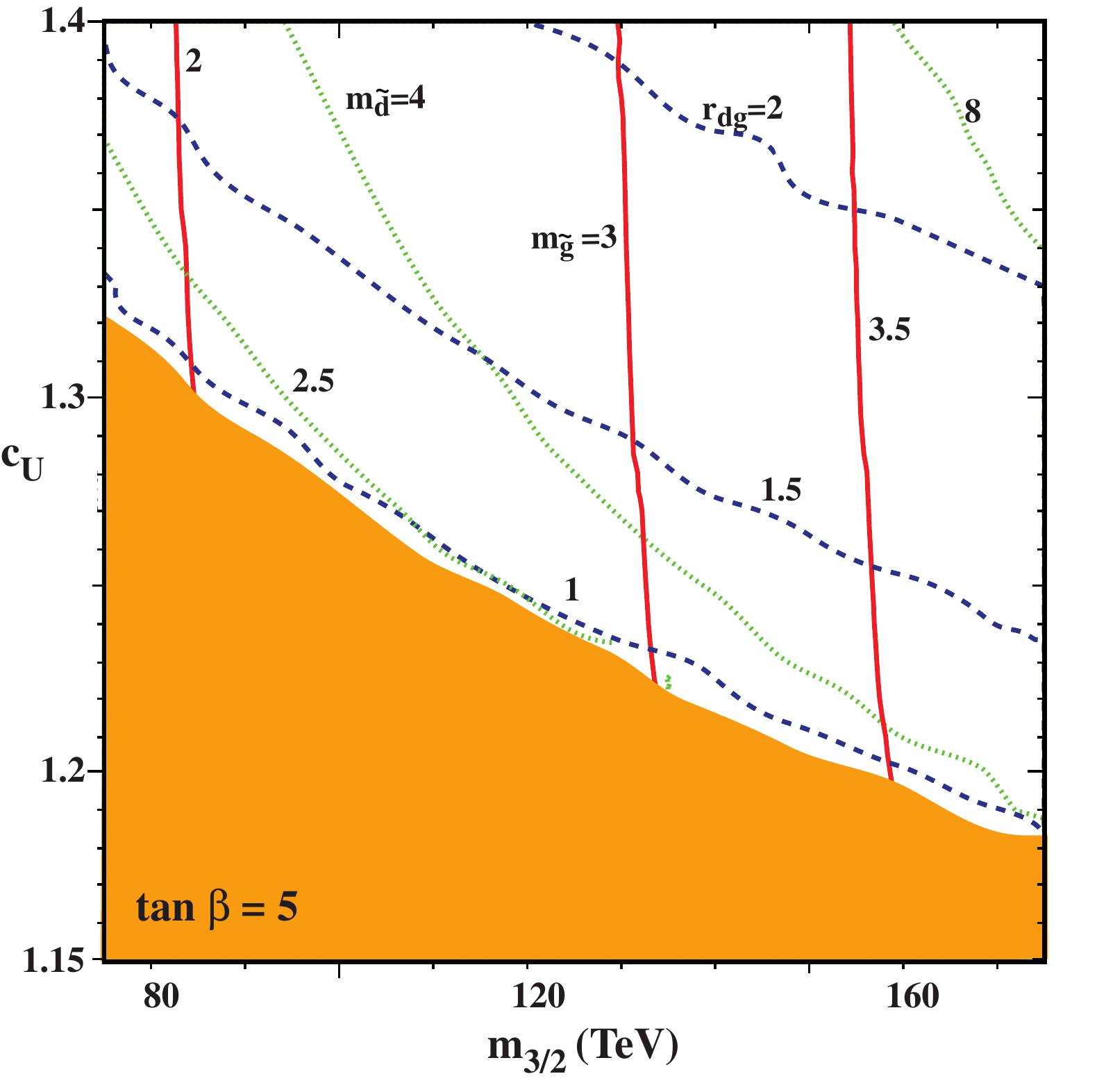}}
\caption{\it We show contours of the sfermion masses in the $m_{3/2}$ versus $c_U$ plane. The contours are as in Fig. (\ref{fig:mhcu}). 
On the right, the solid red line shows the gluino mass contour and the dashed blue line shows the ratio $r_{dg}=m_{\tilde d}/m_{\tilde g}$. The masses are in TeV.\label{fig:m32cu} }
\end{center}
\end{figure}

In Fig. (\ref{fig:m32cu}, we plot the mass contours for $m_{3/2}$ versus $c_U$. As in Fig. \ref{fig:mhcu}, the lower region in the figure is excluded because the scalar down is tachyonic.  As can be seen in this figure it is rather difficult to get the scalar down to be lighter than 2 TeV as $m_{3/2}$ increases. In Fig. (\ref{fig:m32cu}b), we have also plotted the gluino mass as well as the ratio of the down squark mass to the gluino mass, $r_{dg}$. Examining $r_{dg}$ in Fig. (\ref{fig:m32cu}b), we see that the down squark is smaller than the gluino only for regions close to the lower boundary.  These regions correspond to regions where the beta function for the down squark is similar in size to the down squark. This is why this region is somewhat small.  Along this edge we see that the down squark is less than 2.5 TeV only if $m_{3/2}\lesssim 130$ TeV.  This corresponds to a gluino mass of about 3 TeV. By optimizing the parameters we can get down squarks below 2.5 TeV for a gluino mass of about 3.3 TeV.  Regardless, this corresponds to an increased reach in the gluino mass and some interesting prospect for detection at the LHC.

\section{$(g-2)$ of the Muon}
One of the persistent problems facing the SM is the deviation of the SM prediction for $(g-2)_\mu$  with respect to the experimental value. The current deviation in the muon anomalous magnetic moment is \cite{newBNL}
\begin{eqnarray}
\Delta {\rm a}_\mu  =(a_\mu)_{\rm exp} -(a_\mu)_{\rm SM} = (26.1 \pm 8.1) \times 10^{-10}.
\end{eqnarray}
As the LHC has pushed the scale of new physics to higher and higher scales, it is becoming increasingly hard to find explanations for this deviation. In fact, there are few models of supersymmetry which predict a large enough $\Delta {\rm a}_\mu$.

In the mass insertion approximation, the supersymmetric contributions to the anomalous magnetic moment of the muon take the form\footnote{In Eq. (\ref{DelaSU}), $F_1$ is related to
 $\Delta a_\mu^{N1}$, $F_{23}$ is related to $\Delta a_\mu^{N(2-4)}$, and $F_2$ is related to $\Delta a_\mu^{C}$ of \cite{g2muon}.}
\begin{eqnarray}
\Delta {\rm a}_\mu &=& m_\mu^2\tan\beta\mu \left[g_1^2 M_1 F_1(M_1,m_{\tilde \mu_L},m_{\tilde \mu_R})\right.\label{DelaSU}\\
&+&\left. \nonumber g_i^2M_i F_{12}^i(M_i,\mu,m_{\tilde \mu_L},m_{\tilde m_R})+g_2^2M_2 F_2(M_2,\mu,m_{\tilde \nu})\right] \, ,
\end{eqnarray}
where $m_{\tilde \mu_{L,R}}$ are smuon soft masses, and $m_{\tilde \nu}$ are sneutrino soft masses.  For spectra with all SUSY breaking masses and the Higgs bilinear term of similar size, the anomalous magnetic moment of the muon is roughly \cite{g2muon}
\begin{eqnarray}
\Delta {\rm a}_\mu \simeq \frac{1}{32\pi^2} g_2^2\tan\beta \frac{m_\mu}{m_{SUSY}^2} \simeq 2\times 10^{-9} \left(\frac{260 ~{\rm GeV}}{m_{SUSY}}\right)^2\left(\frac{\tan\beta}{10}\right).\label{Dela}
\end{eqnarray}
with the largest contribution coming from $F_2$.   This gives a rough estimate of the size of the Higgs bilinear and slepton masses needed to explain $\Delta {\rm a}_\mu$. In general it is not easy to get sleptons this light while still getting squark masses larger than the LHC constraints.  For this reason it is rather difficult to explain $\Delta {\rm a}_\mu$ in SUSY unless one splits the masses of the first two generations
from that of the third \cite{iyy}.

\begin{figure}
\begin{center}
\includegraphics[width=.7\textwidth]{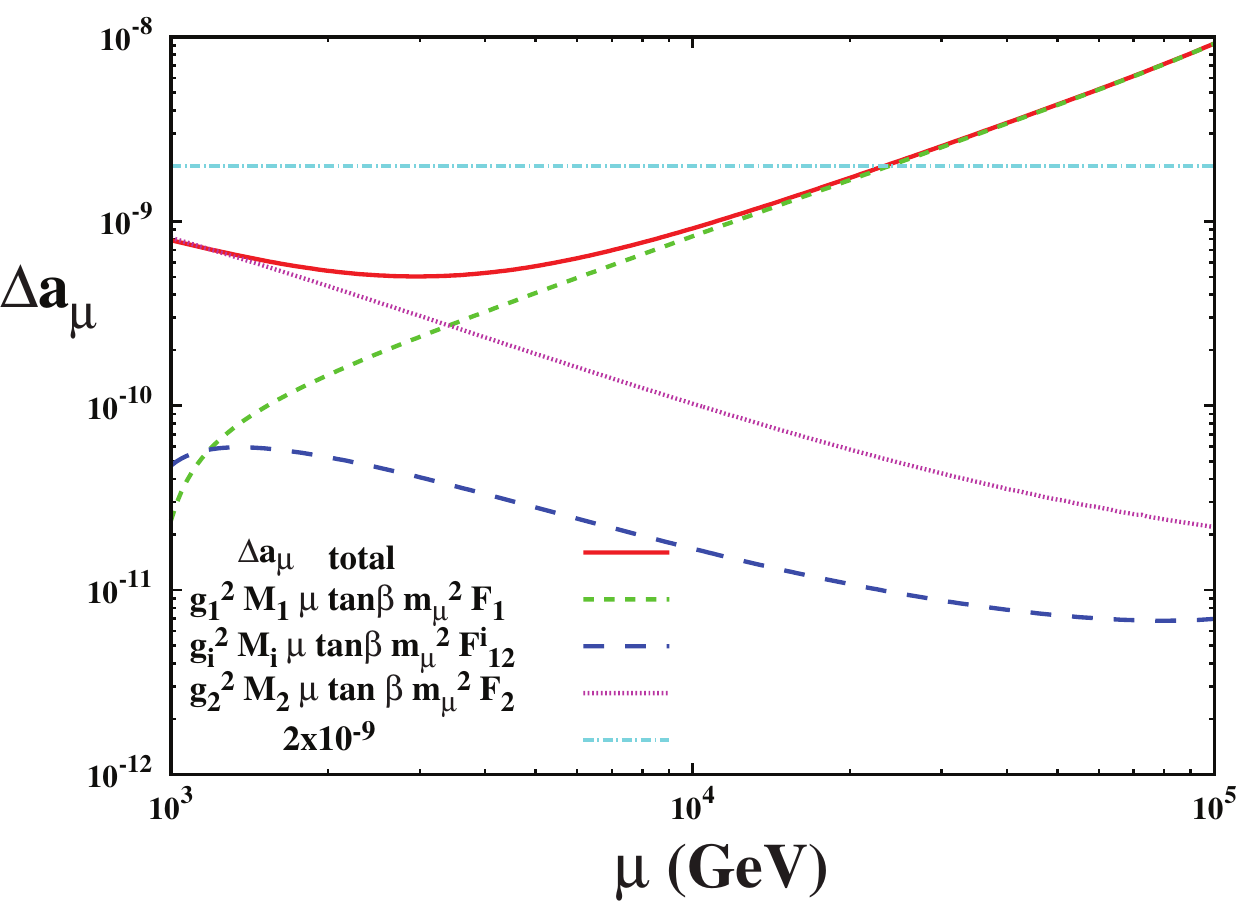}
\caption{\it $\mu$ dependence of the various contributions to $\Delta {\rm a}_\mu$.\label{g2}}
\end{center}
\end{figure}

In PGM, this problem is exacerbated since sfermion masses are pushed to even higher mass scales. Since the size of $\mu$ is related to the stop masses, $\mu$ is also rather large.  If however, the masses of the first two generations are suppressed, $\Delta {\rm a}_\mu$ may increase substantially.  Because $\mu$ is relatively unaffected by this, $F_1$ and $F_{12}^i$ are still suppressed,
\begin{eqnarray}
F_{12}^i\sim F_2\sim \frac{1}{\mu^2m_{\tilde \mu}^2} \, .
\end{eqnarray}
With $F_1$ independent of $\mu$, it has no residual suppression and we have
\begin{eqnarray}
\Delta {\rm a}_\mu = m_\mu^2\tan\beta\mu g_1^2 M_1 F_1(M_i,m_{\tilde \mu_L},m_{\tilde \mu_R}) \, .
\end{eqnarray}
Since this contribution to $\Delta {\rm a}_\mu$ is proportional to $\mu$, it will grow linearly with $\mu$.  To show this important $\mu$ dependence, we have plotted $\Delta {\rm a}_\mu$ with respect to $\mu$ in Fig. (\ref{g2}) for the sample spectrum $M_1=720$ GeV, $M_2=230$ GeV, $m_{\tilde \mu_L}=660$ GeV, and $m_{\tilde \mu_R}=840$ GeV    and $\tan\beta=25$.
With this rather large hierarchy between the first two generation sfermion masses and Higgs bilinear mass, it is possible to explain $\Delta {\rm a}_\mu$ in PGM like models for $\mu\sim m_{3/2} \gtrsim 25$ TeV, even if the smuon masses are larger than $600$ GeV. Fig. (\ref{g2}) also shows the extrapolation between nearly degenerate masses and a hierarchically larger $\mu$. In the region of degenerate masses, the $F_{12}^i$ contribution dominates. For $\mu$ increasing, the $F_1$ quickly becomes the dominant contribution to $(g-2)_\mu$, as we naively argued above.

With heavy third generation masses and light first two generation masses, we also evade another possibly problematic constraint, tachyonic staus.  In PGM, the mixing of the left and right sfermions is proportional to $\mu m_\tau$.  Since the tau mass is non-trivial, the Higgs bilinear mass can not be too much larger than the diagonal soft masses of the stau.  Because the third generation masses are also large in the models we are considering, this constraint is irrelevant.  There is a much weaker constraint coming from having positive masses for the smuons.  However, the muon is much lighter and so these constraints are much weaker.  This much weaker constraint will allow us to push the value of $\mu$ up enough in order to explain $\Delta {\rm a}_\mu$.

\section{Less Simple Unification}
Because of the difficulty in obtaining small sfermion masses in the 1st two generations
with universal constants, $c_U$, we next look at models where $c_1\neq c_2\neq c_3$.  This equates to considering a non-standard breaking of $SU(5)$ or no gauge coupling unification. One possibility for a non-standard breaking of $SU(5)$ is to take the product unification $SU(5)\times U(2)$ or $SU(5)\times U(3)$\cite{PGU}.  In these models, there are three $c_i$. Since the gauge fields of the standard model do not come solely from the $SU(5)$, but are mixtures of the $SU(5)$ gauge field and the additional gauge fields, the PV fields that regulate the low scale gauge fields will be mixed leading to independent $c_i$.  The advantage of considering three $c_i$ is the possibility of light sleptons which can explain the deviation in $(g-2)_\mu$.  Below, we will consider several different scenarios.  Initially, we will scan over generic values of the $c_i$ to see what the parameter space looks like.  Then we will focus on some specific and unique examples which have some interesting results.

\subsection{Generic Coefficients}
In this section, we examine the parameter space for the $c_i$.  As we will see below, the slepton masses are strongly influenced by the Higgs soft masses, $m_{1,2}^2$. For large and negative values of $m_{1,2}^2$, the two-loop gauge running from $SU(2)$ and $U(1)$ are reduced. Since the slepton RG running is independent of $SU(3)$, these will be the dominant contributions to the running making weak scale sleptons easier to realize.

\begin{figure}[t!]
\begin{minipage}{8in}
\epsfig{file=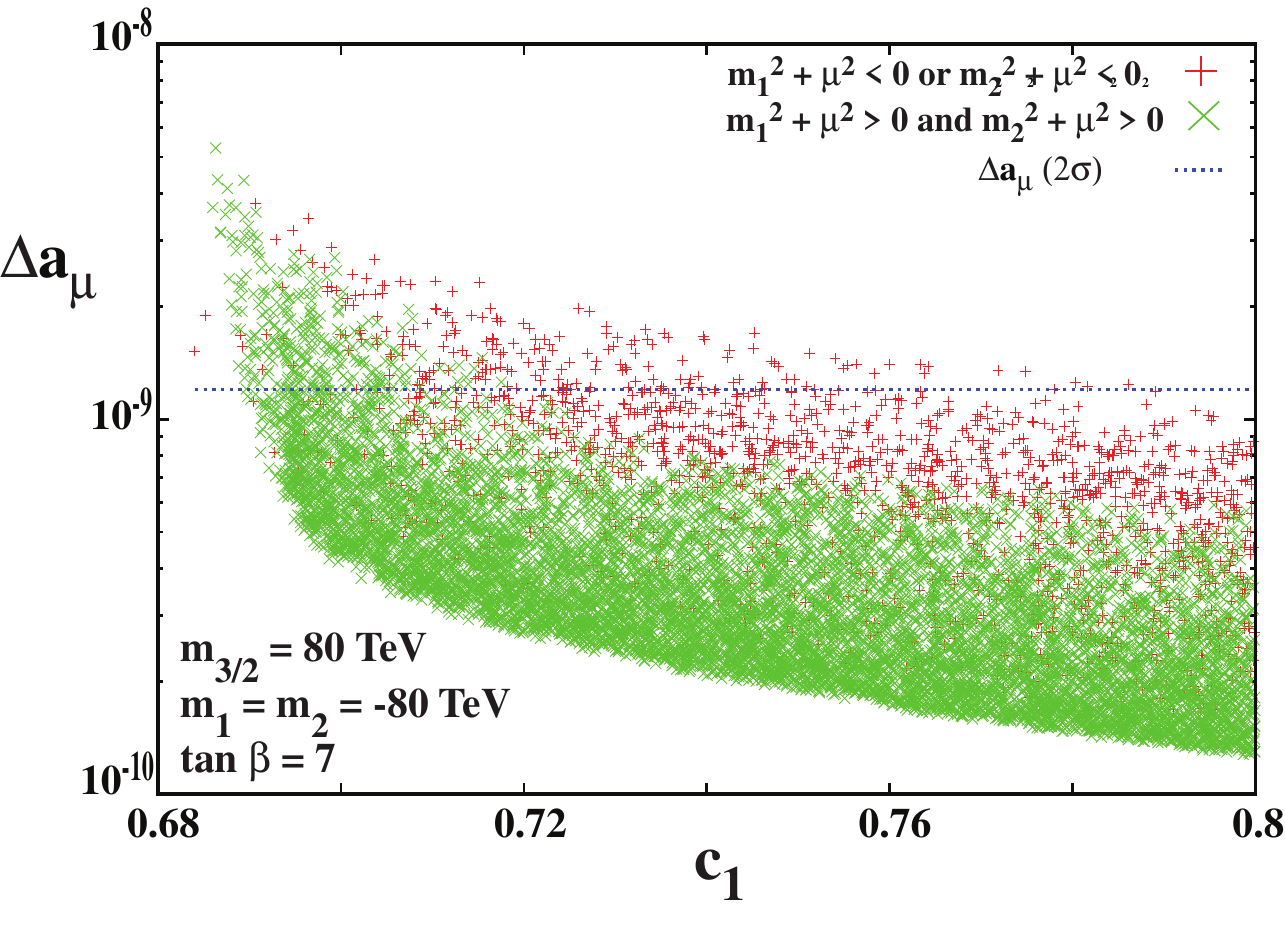,height=2.3in}
\epsfig{file=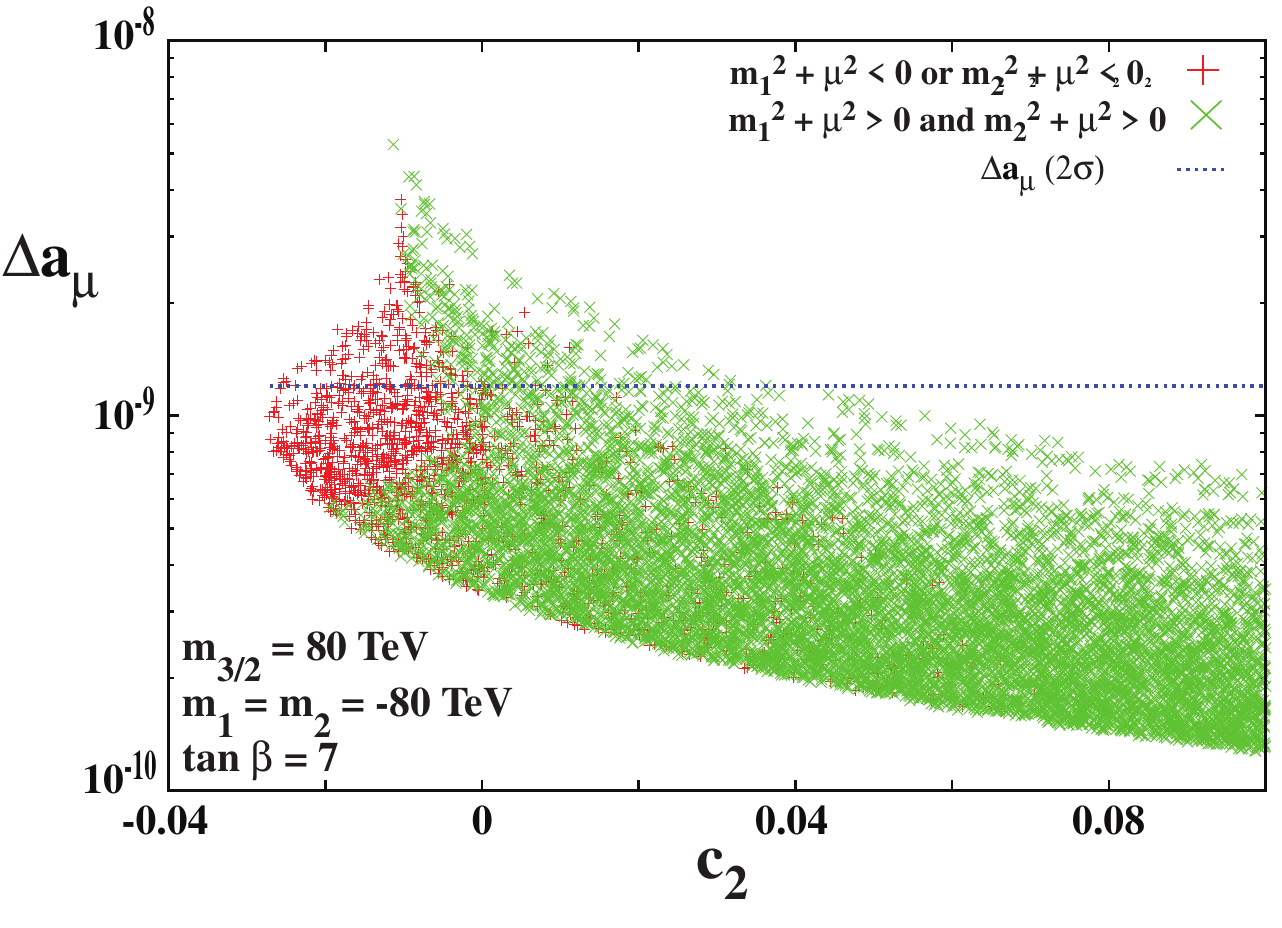,height=2.3in}
\hfill
\end{minipage}
\begin{minipage}{8in}
\epsfig{file=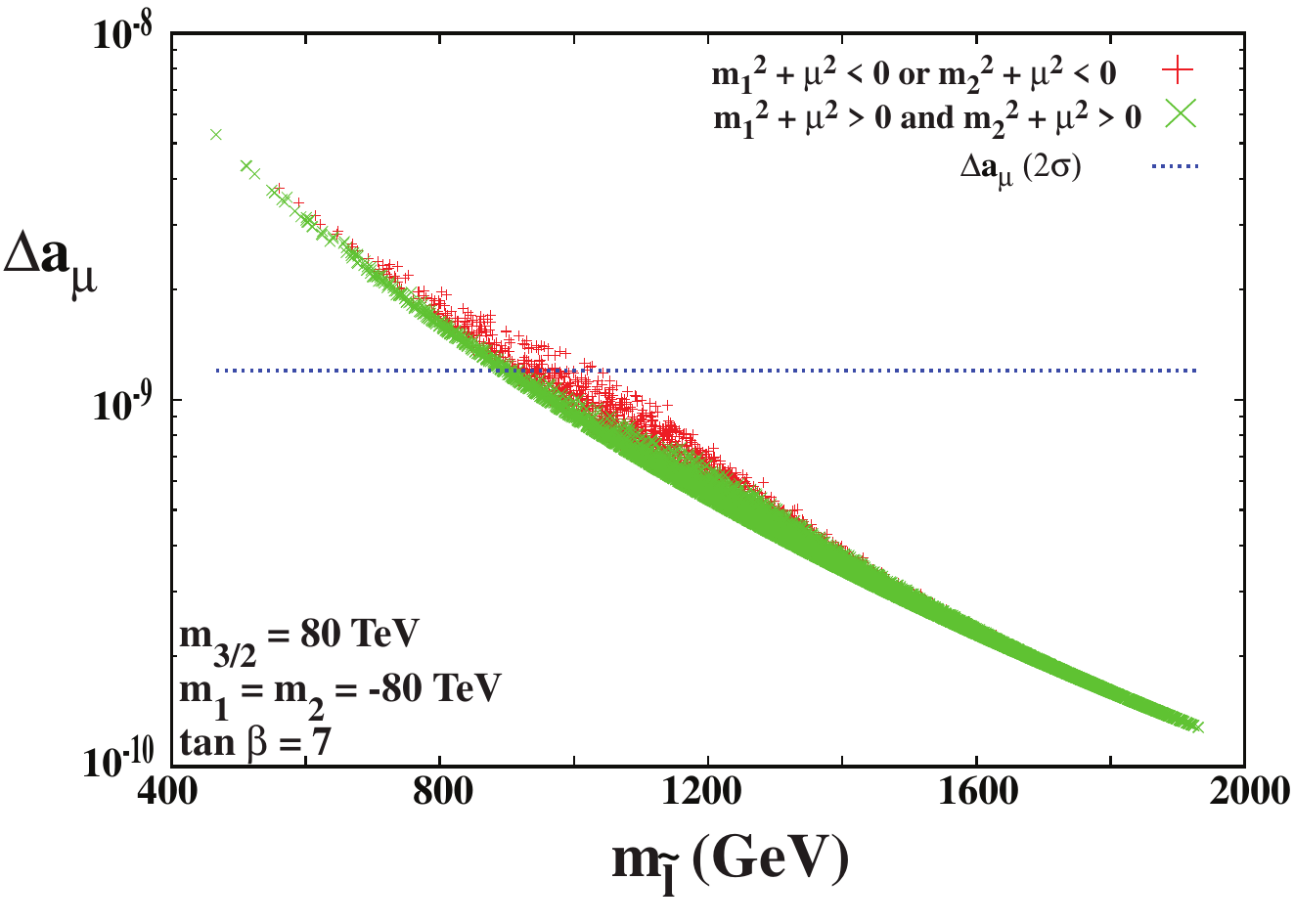,height=2.3in}
\epsfig{file=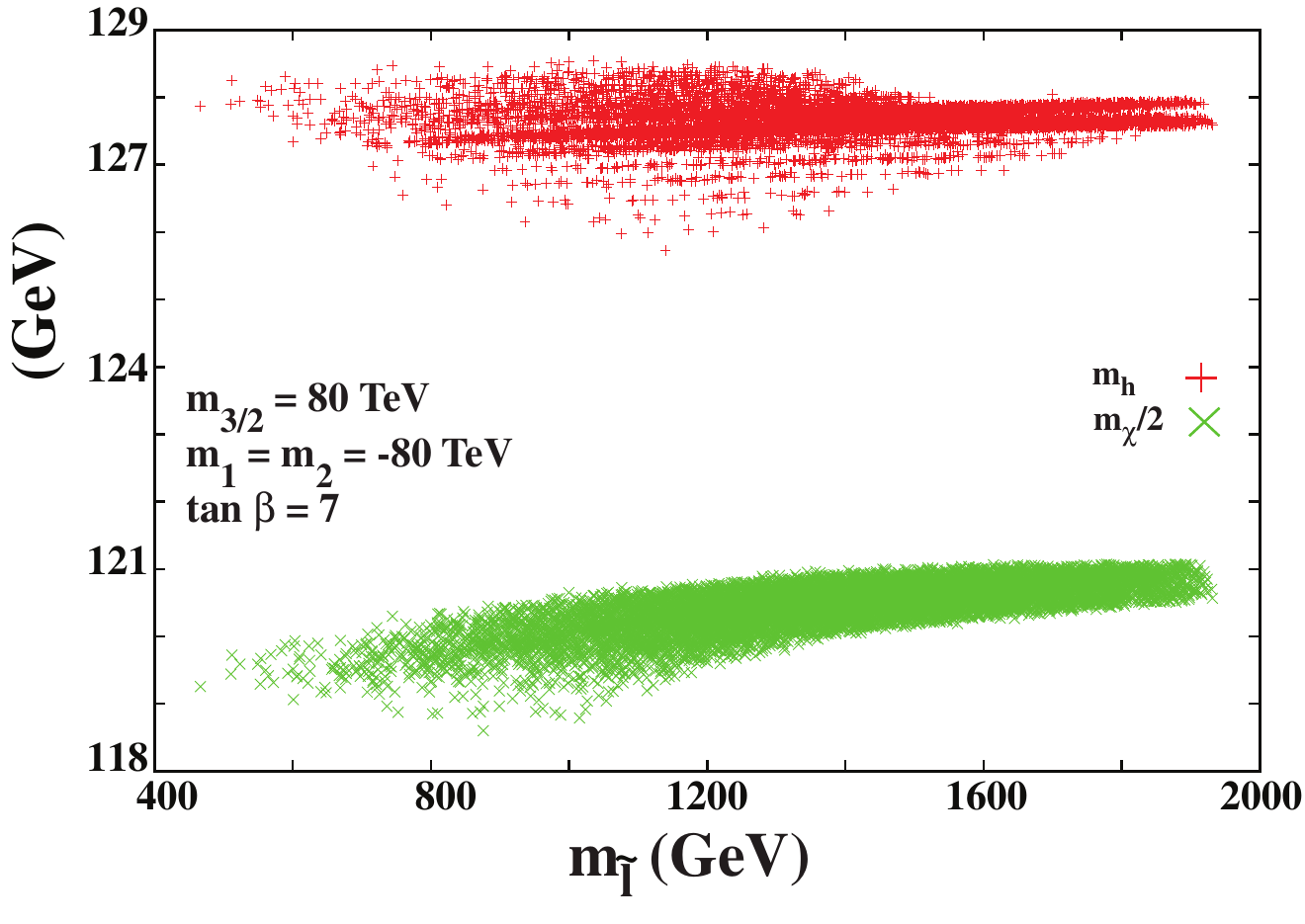,height=2.3in}
\hfill
\end{minipage}
\caption{
{\it The change in the anomalous magnetic moment of the muon, $\Delta {\rm a}_\mu$, with respect to $c_1$ (top left), $c_2$ (top right), and average slepton mass, $m_{\tilde l} \equiv (m_{e_L}+ m_{e_R})/2$ (bottom left) for $m_{1/2}=80$ TeV, $m_1=m_2=-80$ TeV, and $\tan\beta=7$. The dotted line corresponds to the $2\sigma$ lower limit of $\Delta {\rm a}_\mu$. The red +'s have $m_1^2+\mu^2<0$ or $m_2^2+\mu^2<0$ and the green $\times$'s have $m_{1,2}^2+\mu^2>0$.  The bottom right panel  shows the change in the $m_{\chi^0}/2$ (green $\times$'s) and the Higgs mass (red +'s) with respect to the average slepton mass. All four panels are based on the same data.
}}
\label{fig:cgm2}
\end{figure}

With these relations in mind, we examine $m_{3/2}=80$ TeV, $m_1=m_2=-80$ TeV, and $\tan\beta=7$ and scan over the $c_i$. $m_{3/2}=80$ TeV is needed to get a sufficiently large wino mass and $\tan\beta=7$ is chosen so the Higgs mass is sufficiently light.  As mentioned above, $m_1=m_2=-80$ TeV is chosen to reduce the beta functions of the sleptons making it easier to realize weak scale sleptons. We then scan over the $c_i$ with the results found in the top two panels of Fig. (\ref{fig:cgm2}). As can be seen in these figures, the correction to $(g-2)_\mu$ is large enough to account for the experimental discrepancy, but it does require somewhat special values for the $c_i$. In each figure, we distinguish between cases for which one (or both) of the Higgs squared masses is negative at the GUT scale,
$m_i^2 + \mu^2 <0$, for which there are potential cosmological problems \cite{joel}
and those which are always safe since the Higgs squared masses are both always positive.

To better understand the parameter space, we give some additional plots.  In bottom left of Fig. (\ref{fig:cgm2}), we plot the average slepton mass, $m_{\tilde l} = (m_{e_L}=m_{e_R})/2$, with respect to $\Delta {\rm a}_\mu$. As can be seen in these plots, the average slepton mass is rather heavy even for points that can explain $g-2$.  This is due to a large $\mu$.  The other two important parameters for constraining these models, the wino mass (for clarity, $m_\chi/2$ is plotted) and Higgs mass, are also plotted with respect to the average slepton mass in the bottom right figure.  The wino and Higgs mass are fairly independent of the slepton masses. On the other hand, $\Delta {\rm a}_\mu$ is very sensitive to the average slepton mass. Note that
while the Higgs masses shown are somewhat high, we expect that there is a roughly 2 GeV
uncertainty in the calculation of its mass (c.f. \cite{eioy}.

\begin{figure}[t!]
\begin{minipage}{8in}
\epsfig{file=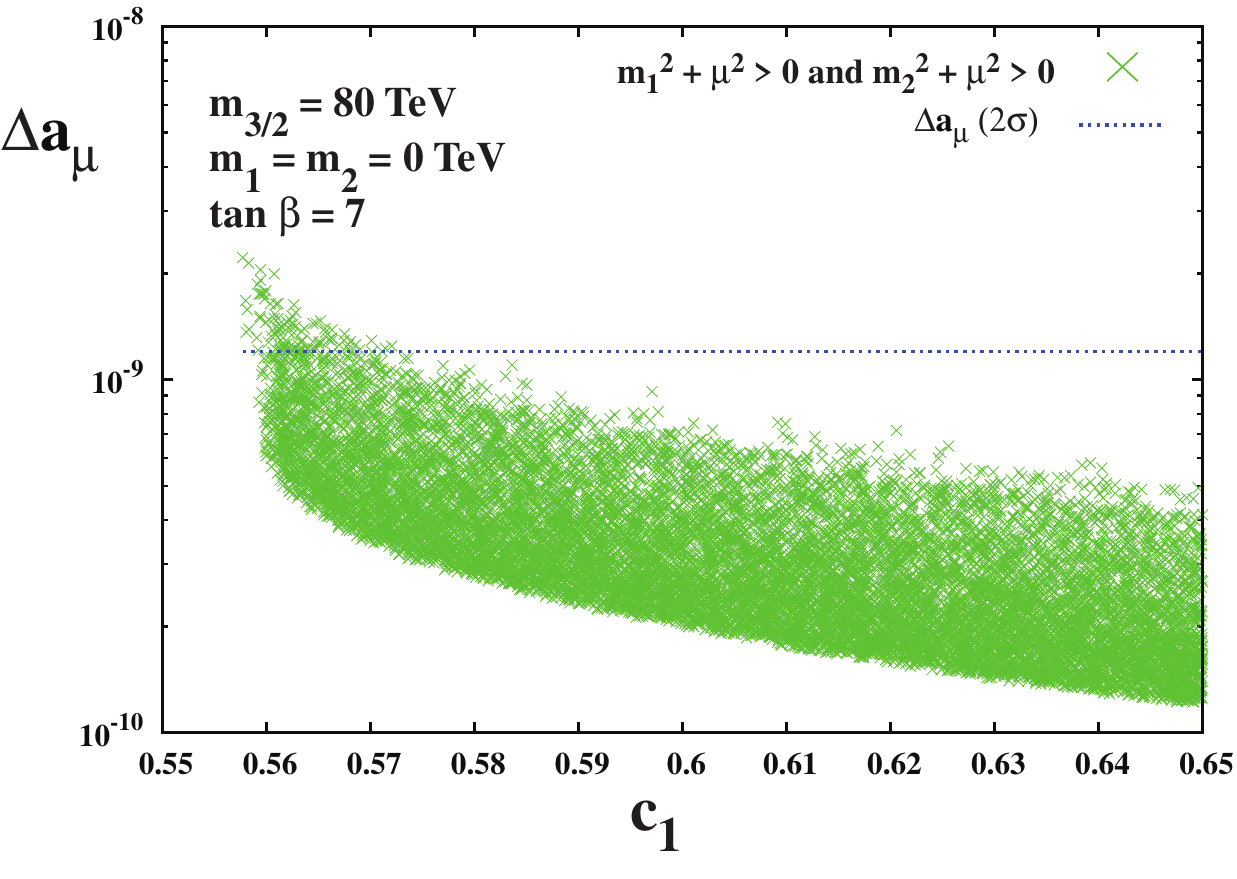,height=2.3in}
\epsfig{file=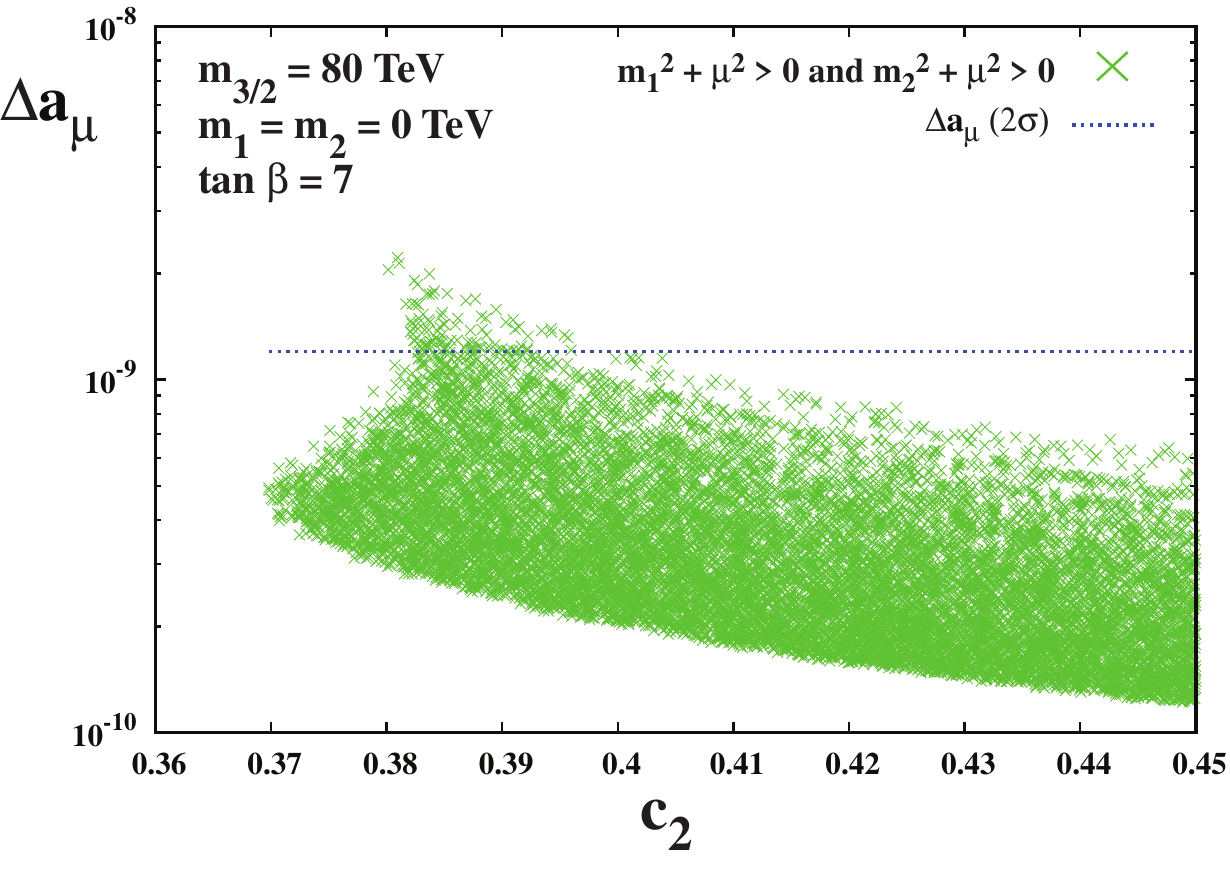,height=2.3in}
\end{minipage}
\begin{minipage}{8in}
\epsfig{file=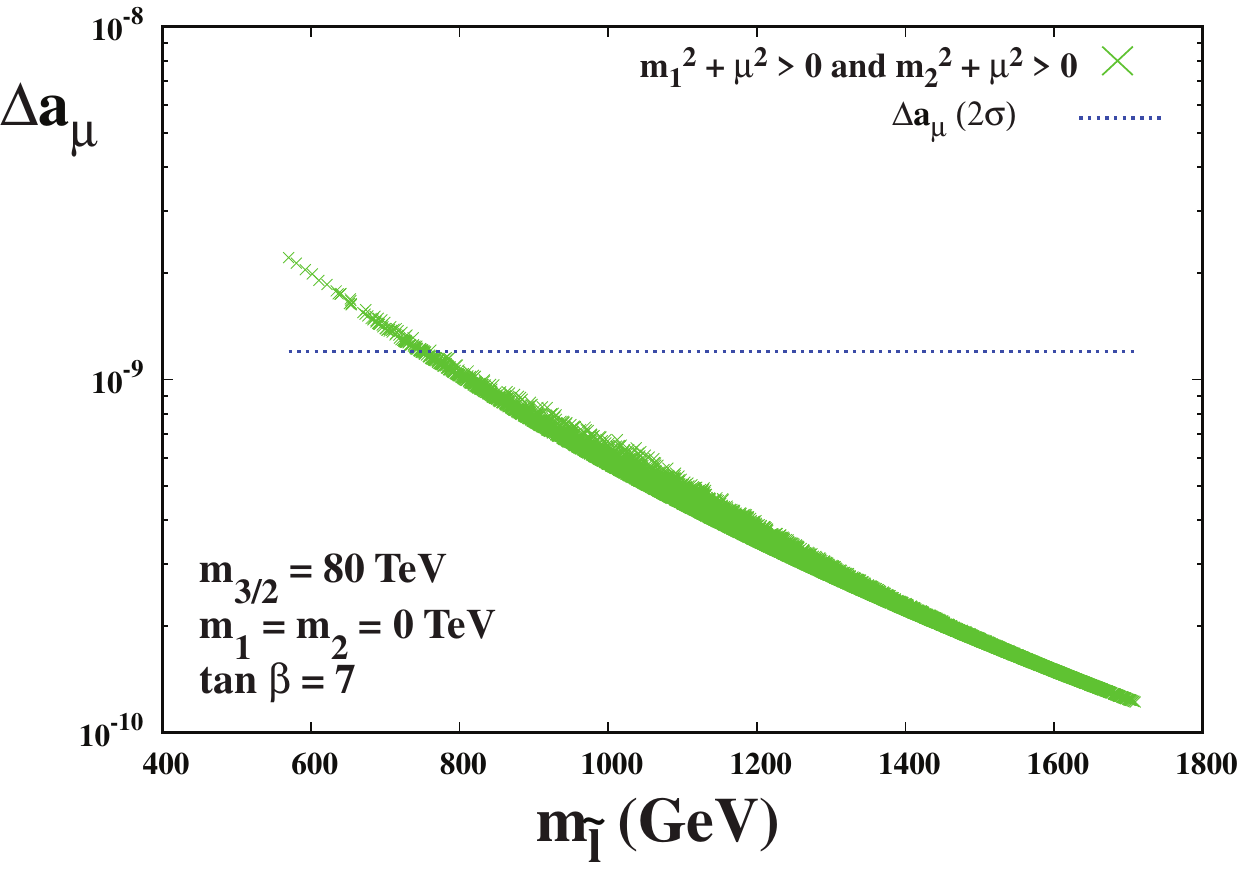,height=2.3in}
\epsfig{file=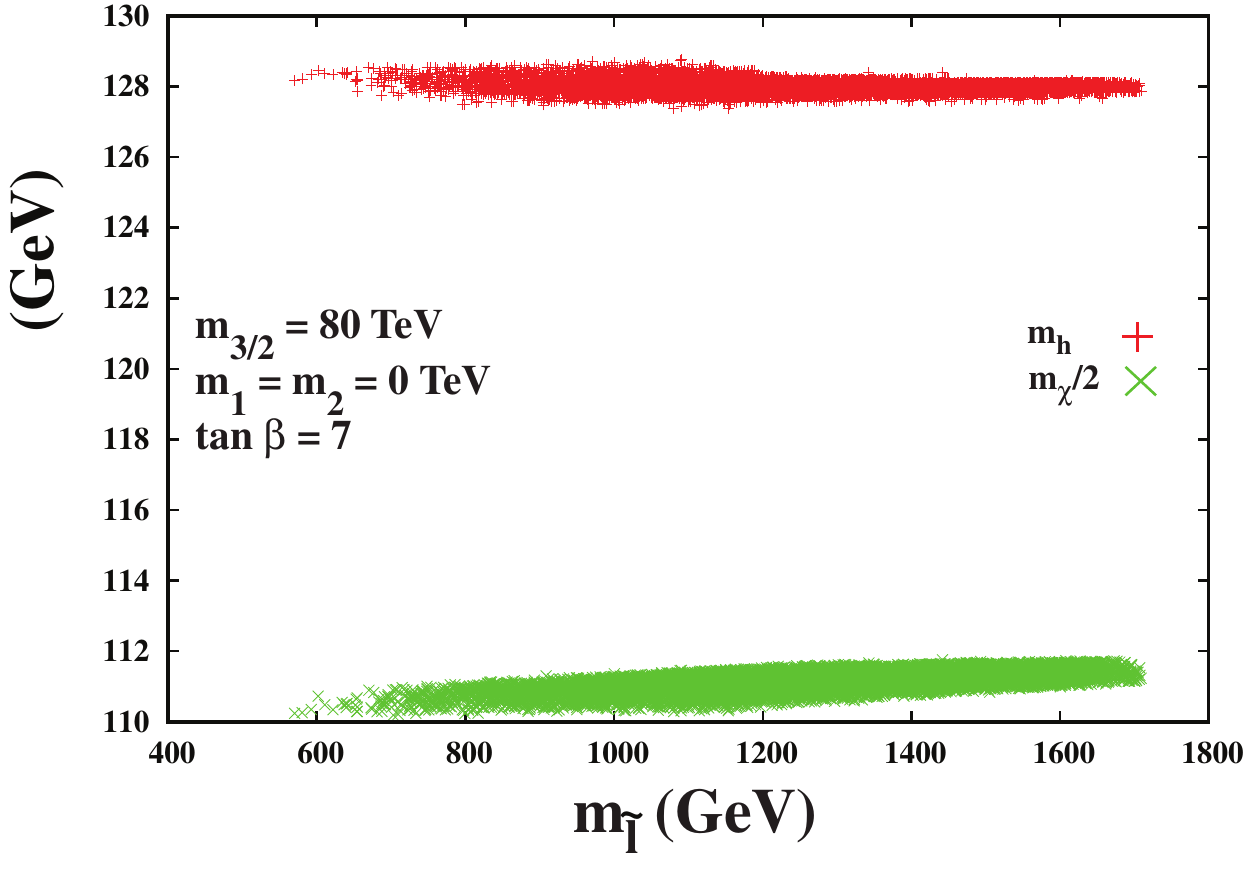,height=2.3in}
\hfill
\end{minipage}
\caption{
{\it Same as Fig. (\ref{fig:cgm2}) except with $m_1=m_2=0$.}}
\label{fig:comp}
\end{figure}

To portray the sensitivity of the RG running on the Higgs soft masses, we show similar plots for $m_1=m_2=0$.  These plots can be seen in Fig. (\ref{fig:comp}).
There are several important things to note.  First, the sleptons tend to have similar sizes since this is predominantly set by $m_{3/2}$.  However, the lighter slepton masses arise for $c_i$ which are tuned to a greater degree. Another important difference is a large decrease in the wino mass.  This is due to a significant change in $m_A$ and $\mu$.  Because the threshold corrections to the wino depend strongly on both $\mu$ and $m_A$, the wino mass is much lighter for $m_1=m_2=0$. This is an additional reason why $m_1=m_2=-80$ TeV is advantageous. For $m_1=m_2=0$, we would need to take a larger value of $m_{3/2}$ making it more difficult to get weak scale sleptons.

\subsection{General Coefficients with Light Squarks}
\begin{figure}
\begin{minipage}{8in}
\epsfig{file=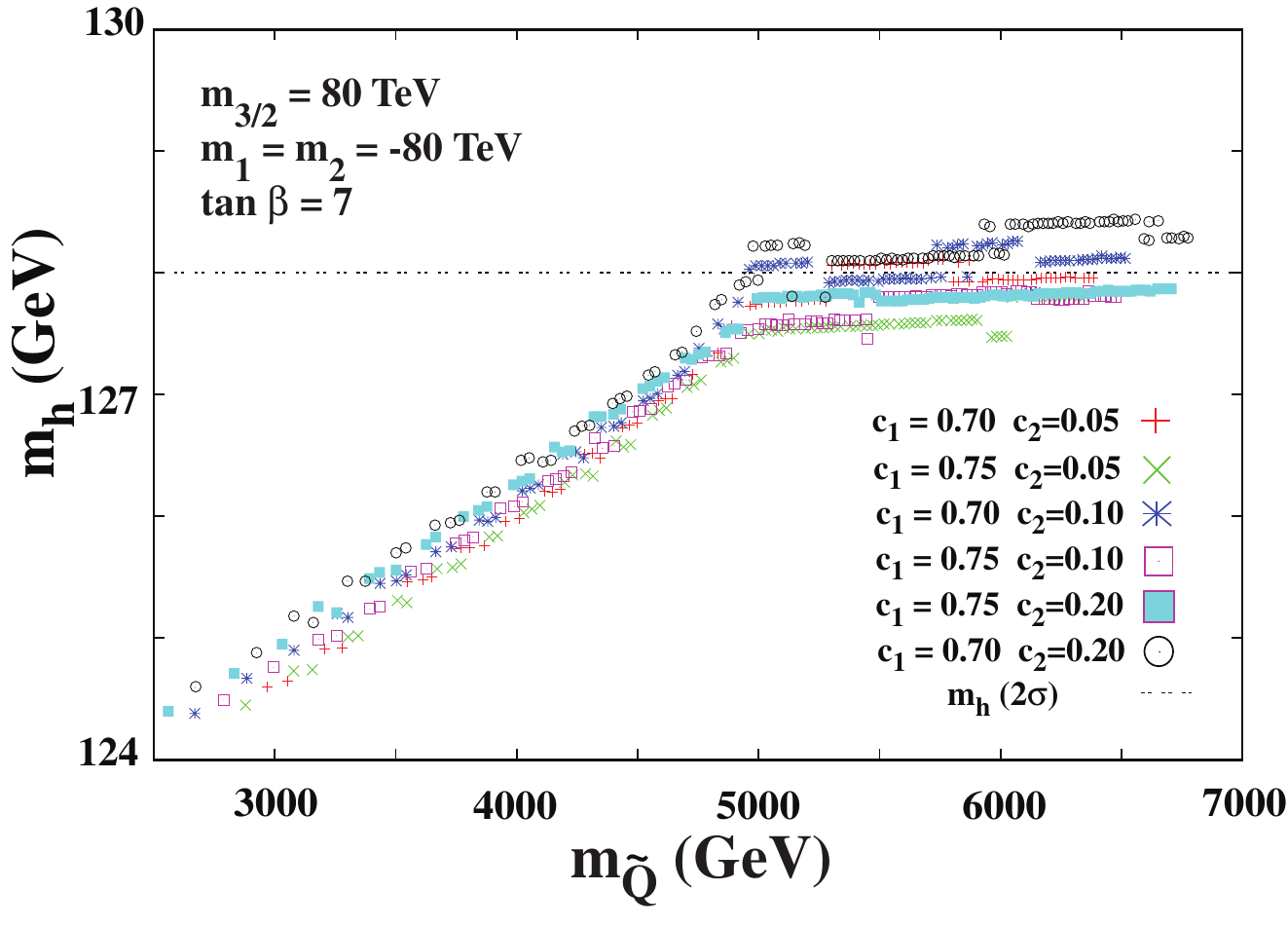,height=2.3in}
\epsfig{file=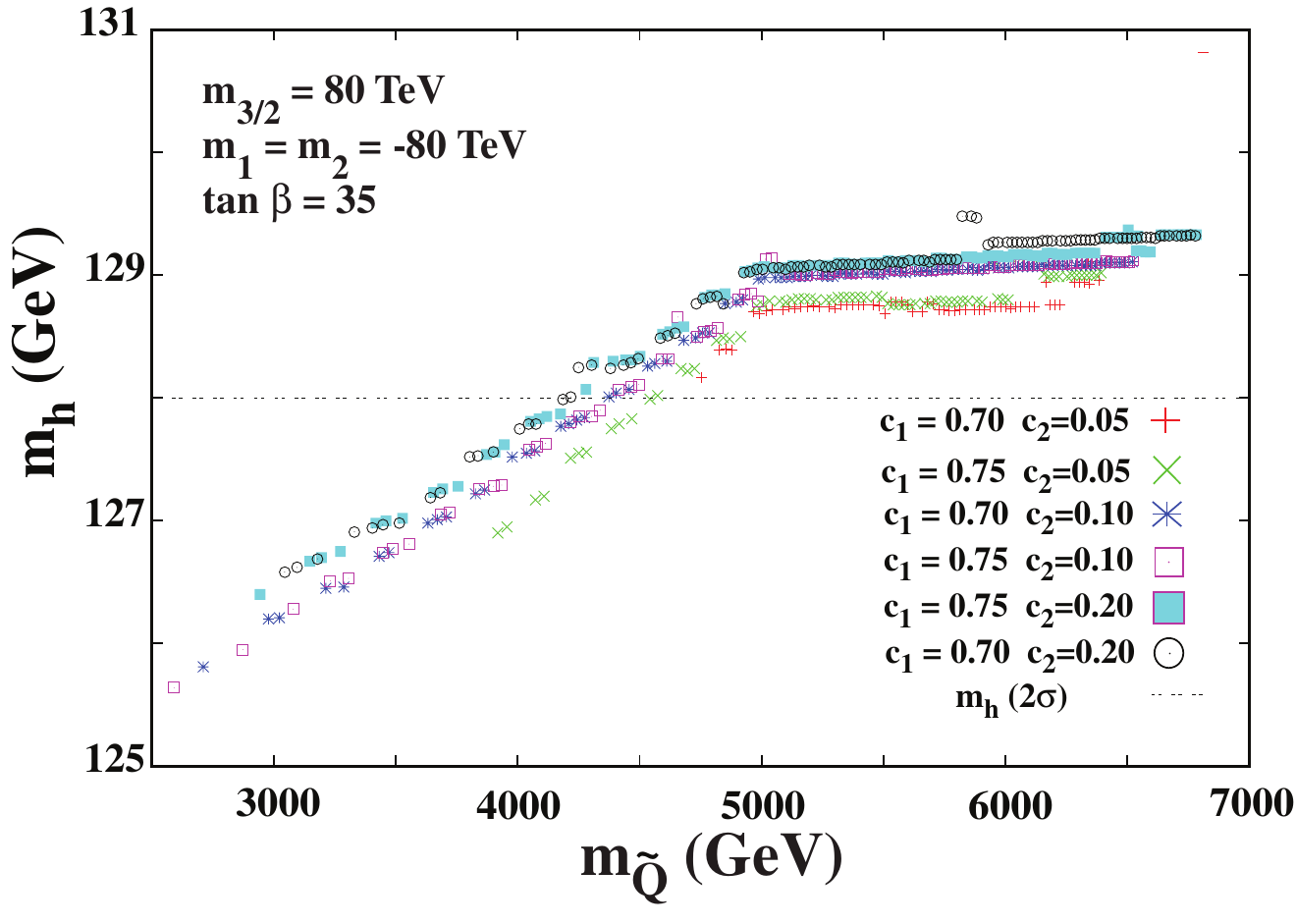,height=2.3in}
\hfill
\end{minipage}
\caption{\it Here we plot the Higgs mass versus the mass of the left-handed squark mass of the first two generations for (left) $\tan\beta=7$  and (right) $\tan\beta = 35$. The red +'s are for $c_1=0.7$ and $c_2=0.05$. The green $\times$'s are for $c_1=0.75$ and $c_2=0.05$. The blue stars are for $c_1=0.7$ and $c_2=0.1$. The magenta boxes are for $c_1=0.75$ and $c_2=0.1$. The cyan filled boxes are for $c_1=0.7$ and $c_2=0.2$. The gray circles are for $c_1=0.7$ and $c_2=0.2$. The horizontal dashed line corresponds to the 2 $\sigma$ lower limit on $\Delta {\rm a}_\mu$.  \label{fig:higmas} }
\vspace{.2in}
\end{figure}

Next, we examine some special values of the $c_i$ which tend to be interesting. In particular, we first allow $c_3$ to vary so that we obtain light first and second generation squarks. Although it may seem this has no affect on $g-2$, it will have some rather important and unexpected effects. In Fig. (\ref{fig:higmas}), we examine the Higgs mass for different but fixed values of $c_{1,2}$ and vary $c_3$ which we will parameterize by the left-handed squark mass, $m_{\tilde Q}$.

\begin{figure}
\begin{minipage}{8in}
\epsfig{file=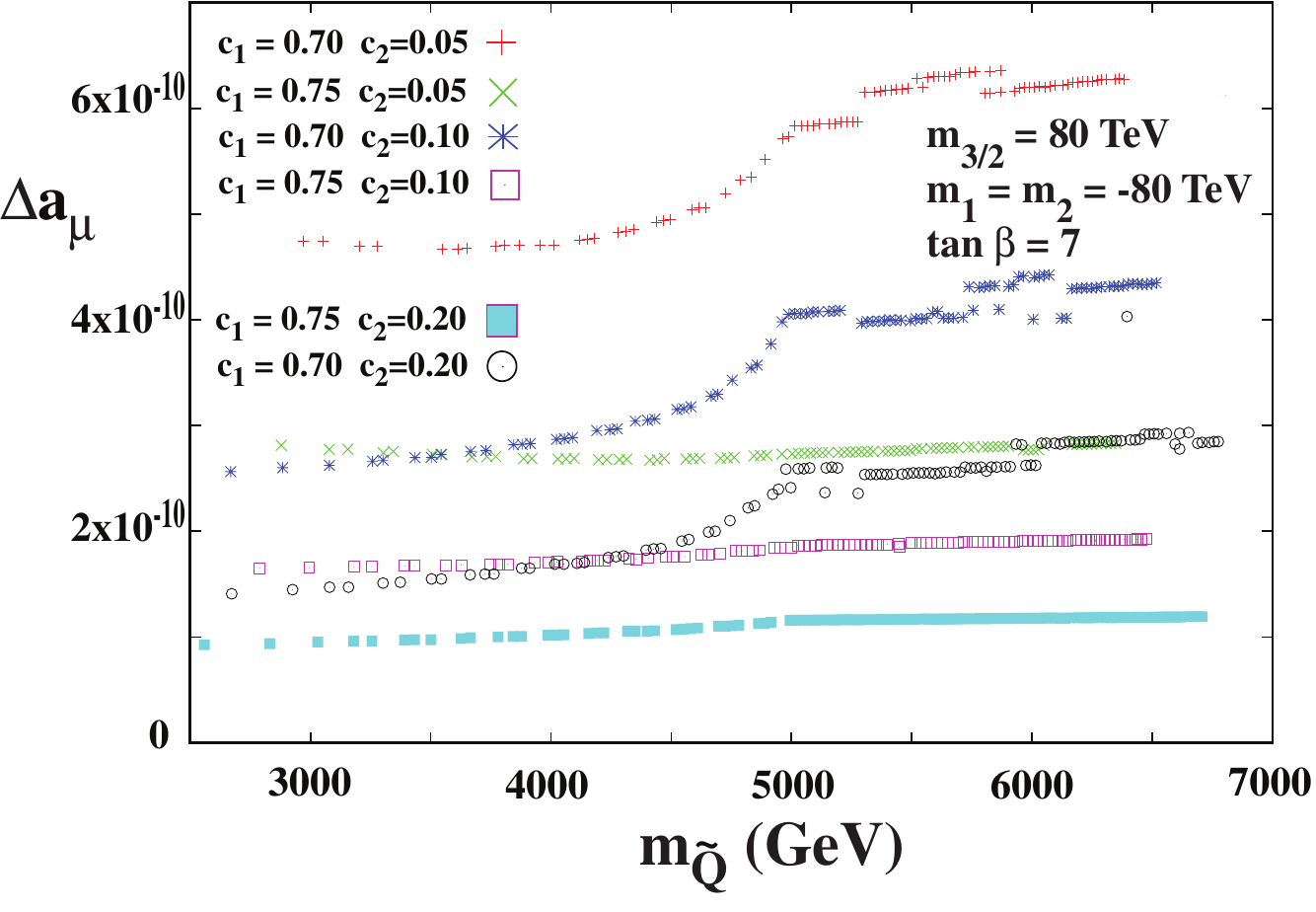,height=2.3in}
\epsfig{file=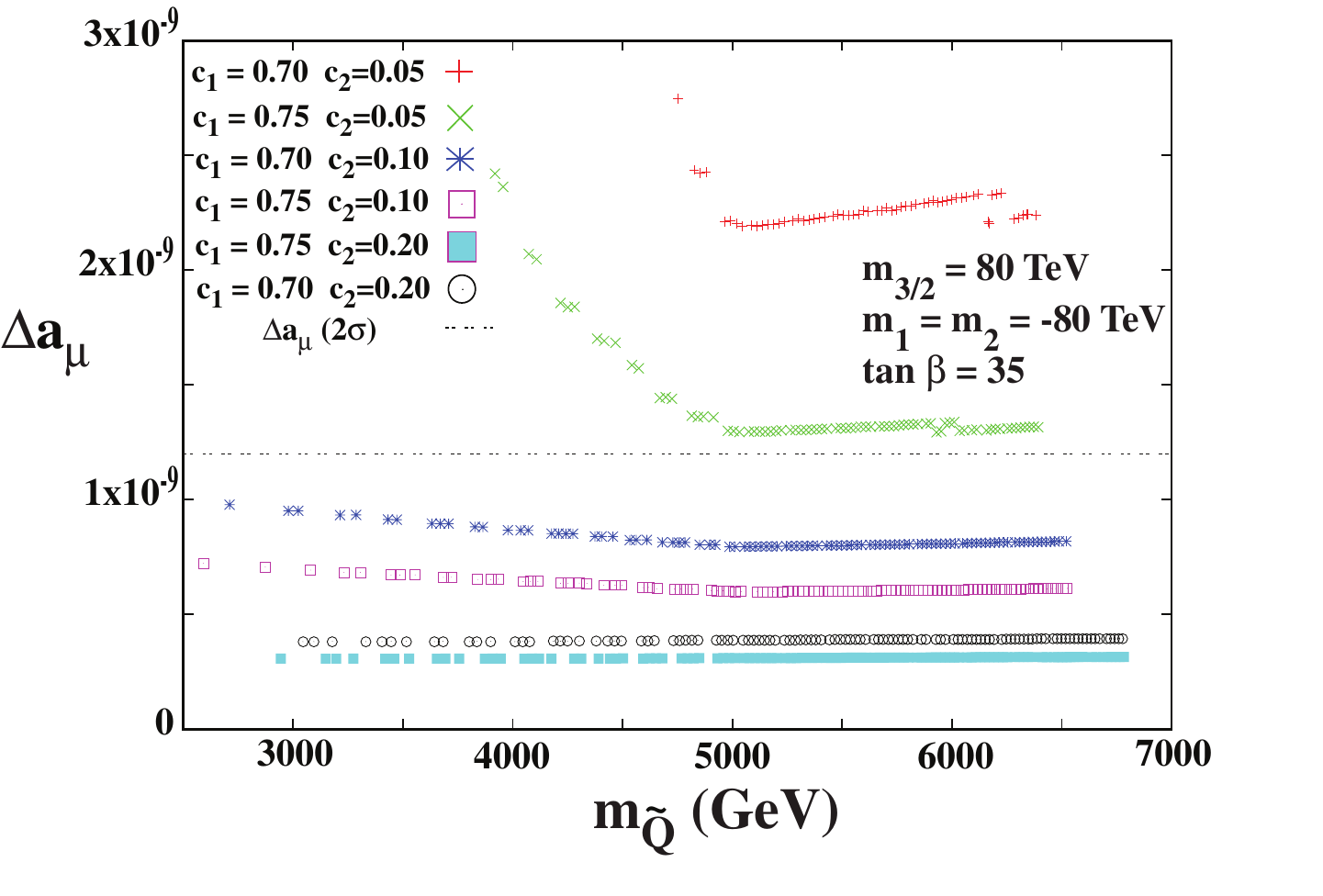,height=2.3in}
\hfill
\end{minipage}
\caption{\it Here we plot the change in the anomalous magnetic moment $\Delta {\rm a}_\mu$ with respect the left-handed squark mass for (left) $\tan\beta=7$ and (right) $\tan\beta =35$ . The symbols used are identical to that in Fig. (\ref{fig:higmas}). \label{fig:gm2msq} }
\end{figure}

In this figure, we see that as soon as $m_{\tilde Q} \lesssim 5$ TeV, the Higgs masses begins to decrease, although naively, it would be expected that the Higgs mass is independent of the mass of first two generation squarks. This behavior is important for explaining the deviation in $(g-2)_\mu$, because it allows us to push up the value of $\tan\beta$ and still have a sufficiently small Higgs mass. The Higgs mass is sensitive to the first two generation squark masses through alterations in the running of the gauge and Yukawa couplings. When the first, second, and third generation sfermion masses are similar there are effectively two regions of RG running, above and below the sfermion mass scale.  However, if the first and second generations are sufficiently separated from the third generation, there is a third region that emerges.  In this third region, the beta function for $SU(2)$ nearly vanishes.  This leads to rather large deviations in the gauge couplings for the scale where the third generation decouples. This deviation in the coupling leads to a significant change in the Higgs mass. As it turns out, we can get a light enough Higgs mass even for large $\tan\beta$. Because of this new found freedom in $\tan\beta$, we can further enhance $\Delta {\rm a}_\mu$ in the region where the squark masses are light by taking $\tan\beta$ large.  This enhancement of $\Delta {\rm a}_\mu$ for regions with light squark masses can be seen in Fig. (\ref{fig:gm2msq}).

\begin{figure}
\begin{minipage}{8in}
\epsfig{file=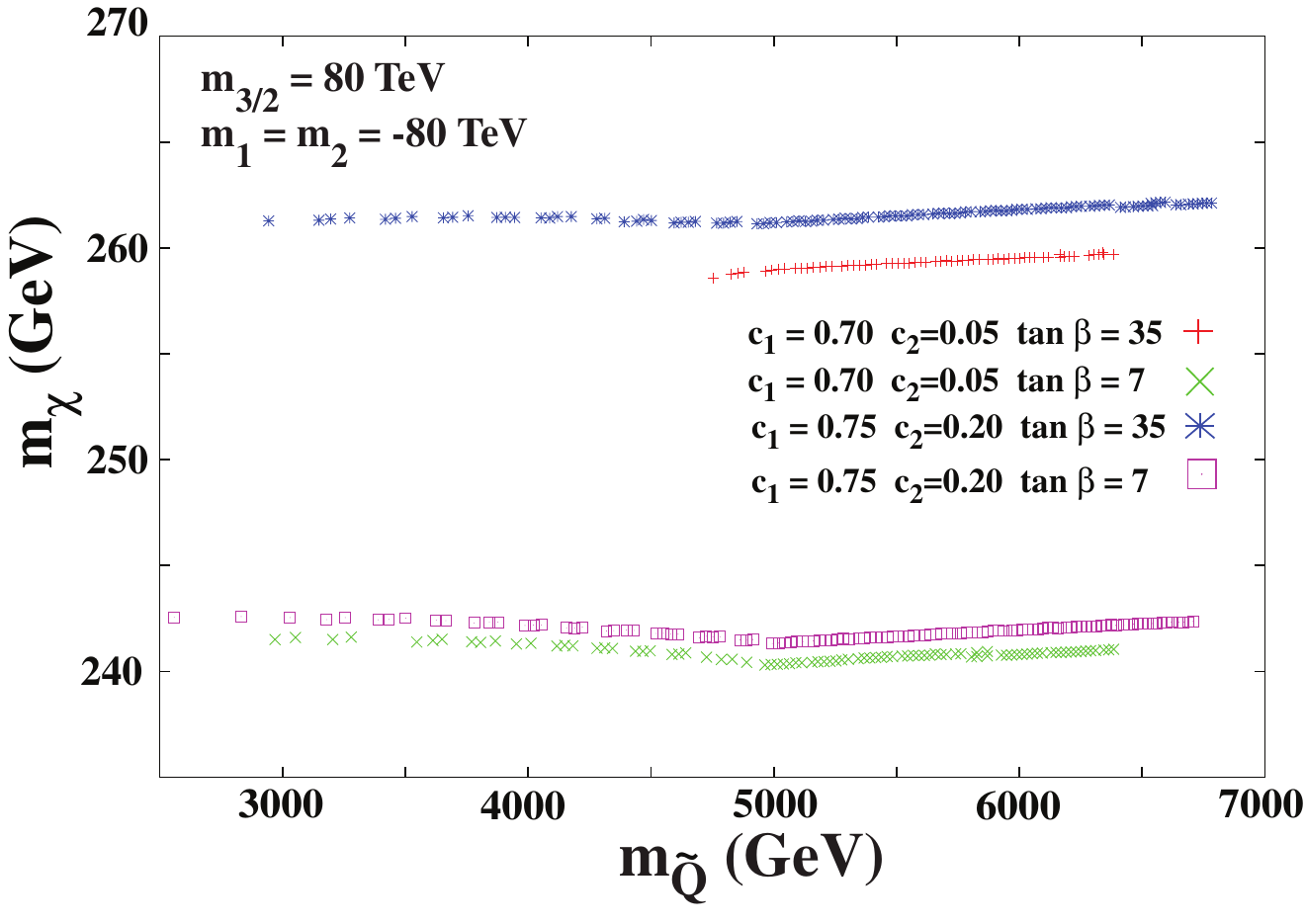,height=2.3in}
\epsfig{file=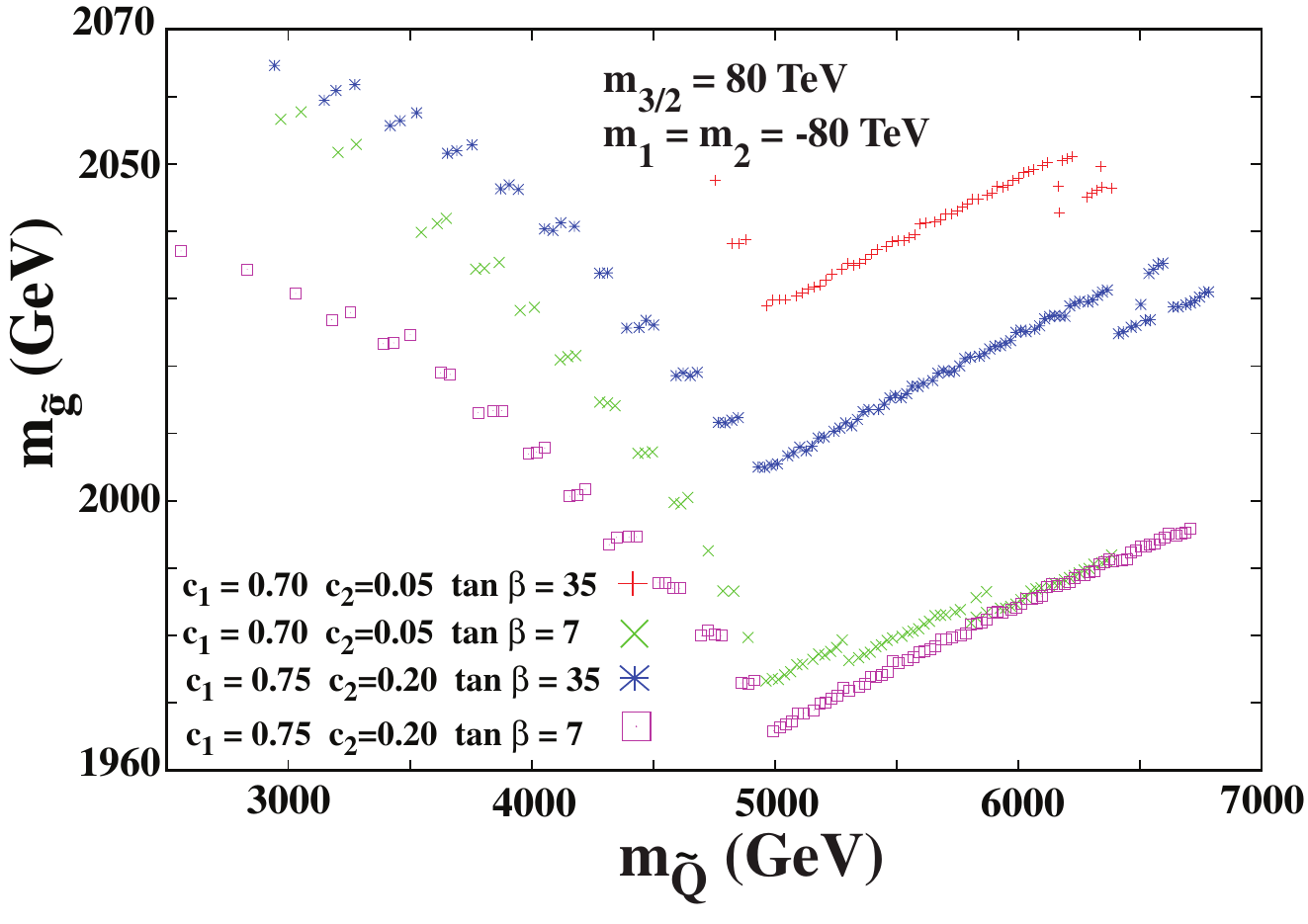,height=2.3in}
\hfill
\end{minipage}
\caption{\it Here we plot the change in the neutralino mass, $m_{\chi^0}$, with respect to the left-handed squark mass. The red +'s are for $c_1=0.7$ and $c_2=0.05$ with $\tan\beta=35$. The green $\times$'s are likewise for $\tan\beta=7$. The blue stars are for $c_1=0.75$ and $c_2=0.2$ with $\tan\beta=35$. The magenta boxes are likewise for $\tan\beta=7$. \label{fig:inomsq} }
\end{figure}

\begin{figure}
\begin{minipage}{8in}
\epsfig{file=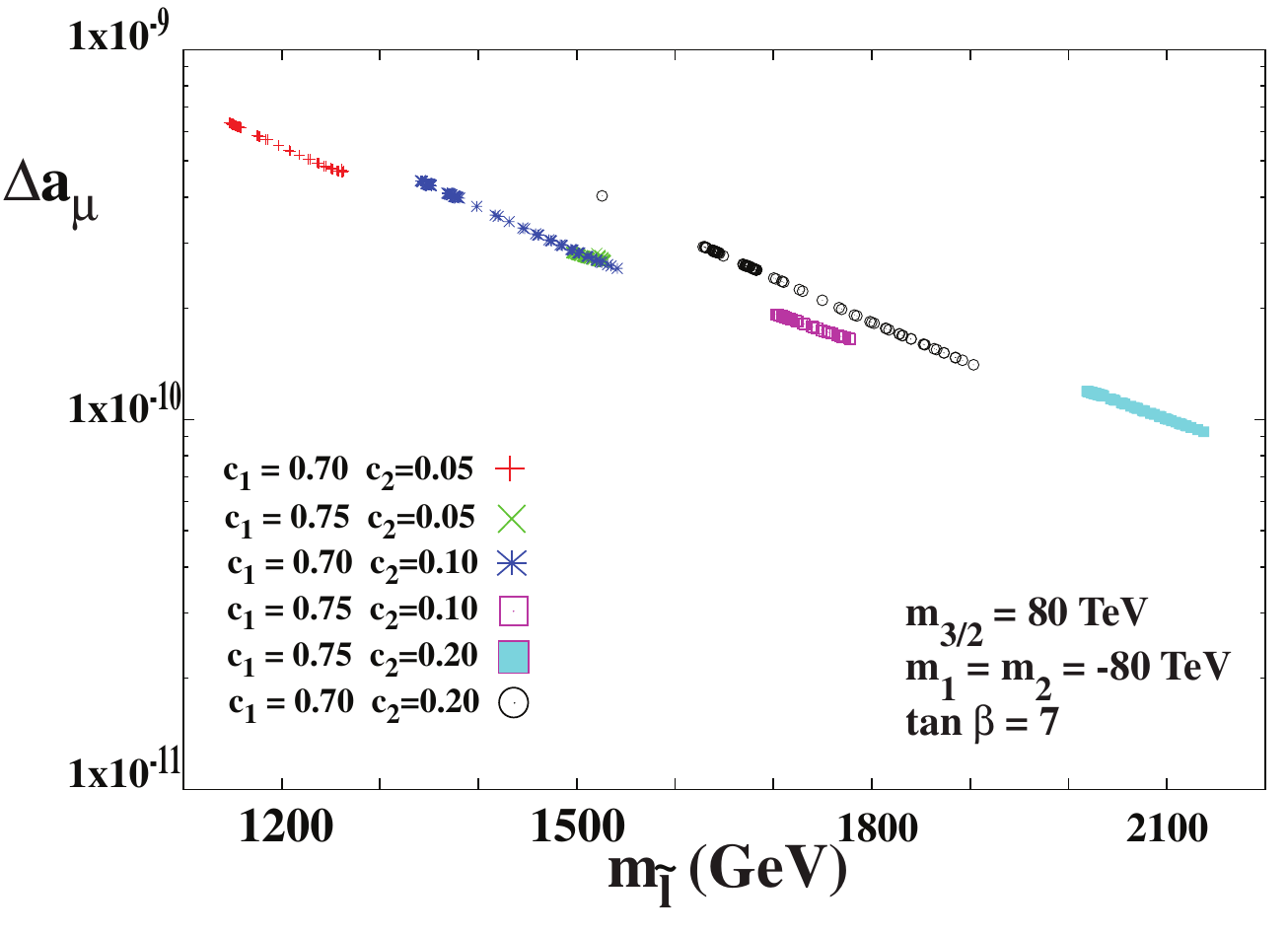,height=2.3in}
\epsfig{file=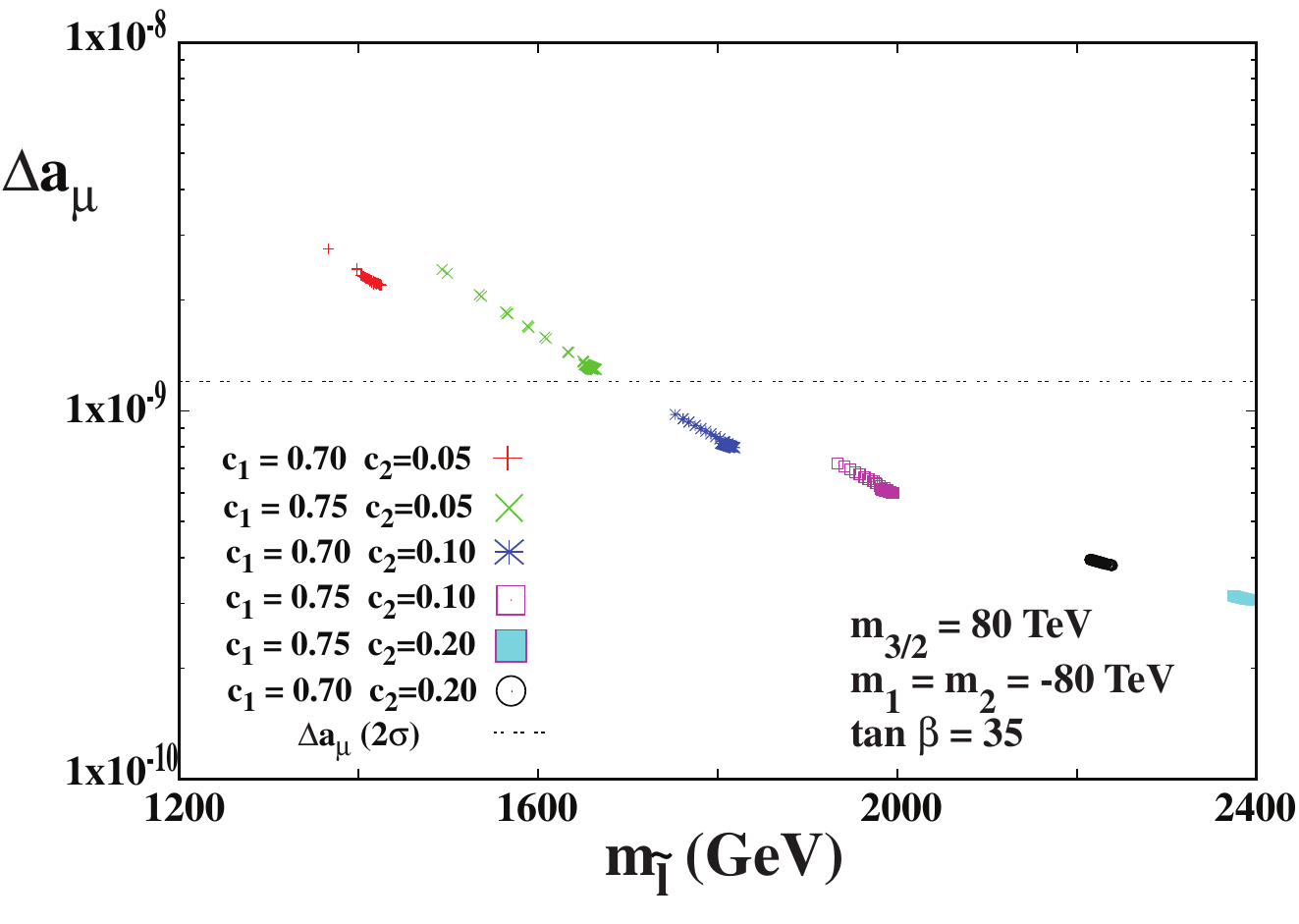,height=2.3in}
\hfill
\end{minipage}
\caption{\it Here we plot the change in the anomalous magnetic moment with respect to the average slepton masses for (left) $\tan\beta=7$ and (right) $\tan\beta =35$ . The symbols used are identical to that in Fig. (\ref{fig:higmas}).
\label{fig:amumel} }
\end{figure}

Since the gauge couplings are deflected by the alteration of the beta functions from light first and second generation sfermions, we will also see changes in the masses of the gauginos.  These changes are fairly mild as can be seen in Fig (\ref{fig:inomsq}), although the scaling on the axis makes it appear somewhat drastic.
For completeness, we also plot the slepton masses versus the anomalous magnetic moment.  This is shown in Fig. (\ref{fig:amumel}).

\subsection{General Coefficients Plus....}
Finally, we consider some models which can relax the constraint on the wino mass. Since it is this constraint which is responsible for pushing up the gravitino mass, relaxing this constraint will drastically reduce the fine tuning needed to get light sleptons. There are actually two simple ways to evade the wino mass constraint: increase its mass for a given gravitino mass or change the decay width of the wino. Both of these mechanisms require dark matter to come from some source other  than the lightest supersymmetric particle (LSP). However, since the constraints on wino dark matter are getting ever more stringent \cite{wino}, it is worth examining the case where the wino is not the dominant source of dark matter. One interesting possibility is to assume that dark matter arises from a PQ like theory. The fields responsible for PQ symmetry breaking then act as messenger for the gauginos\cite{axmed} enhancing the gaugino masses for a given value of the gravitino mass. Another option is to allow $R$-parity violation. This relaxes the constraint on the wino by increasing its decay width.  In PGM, the LSP is a neutral wino and the charged wino is about $160$ MeV heavier. Because these particles are nearly degenerate there is a strong phase space suppression of the decay. If R-parity violating interactions are included, the charged wino can decay directly to standard model particles alleviating the phase space suppression\footnote{To evade baryon asymmetry washout, some model building is needed. See the review\cite{RPV} }.

\begin{figure}
\begin{minipage}{8in}
\epsfig{file=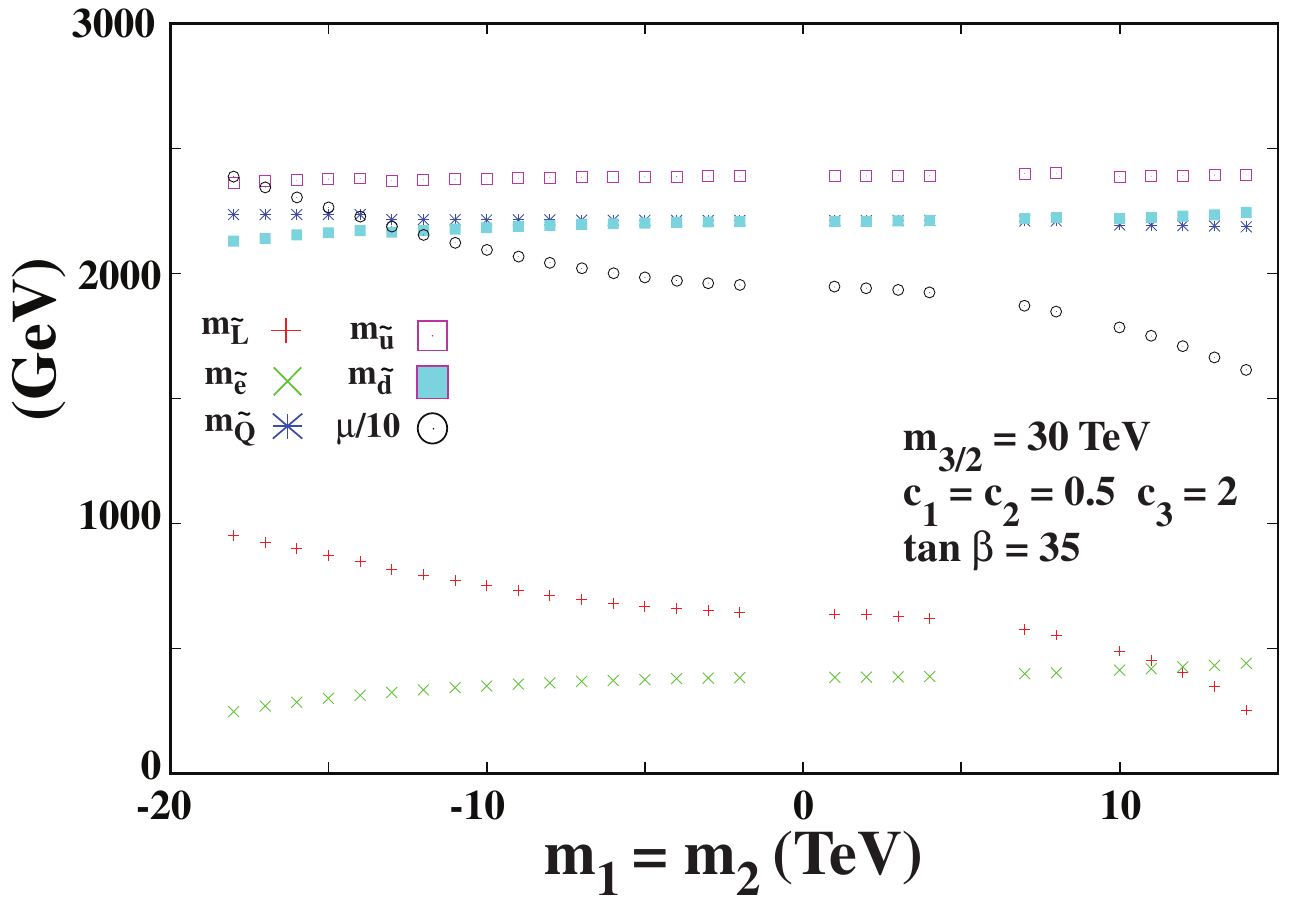,height=2.3in}
\epsfig{file=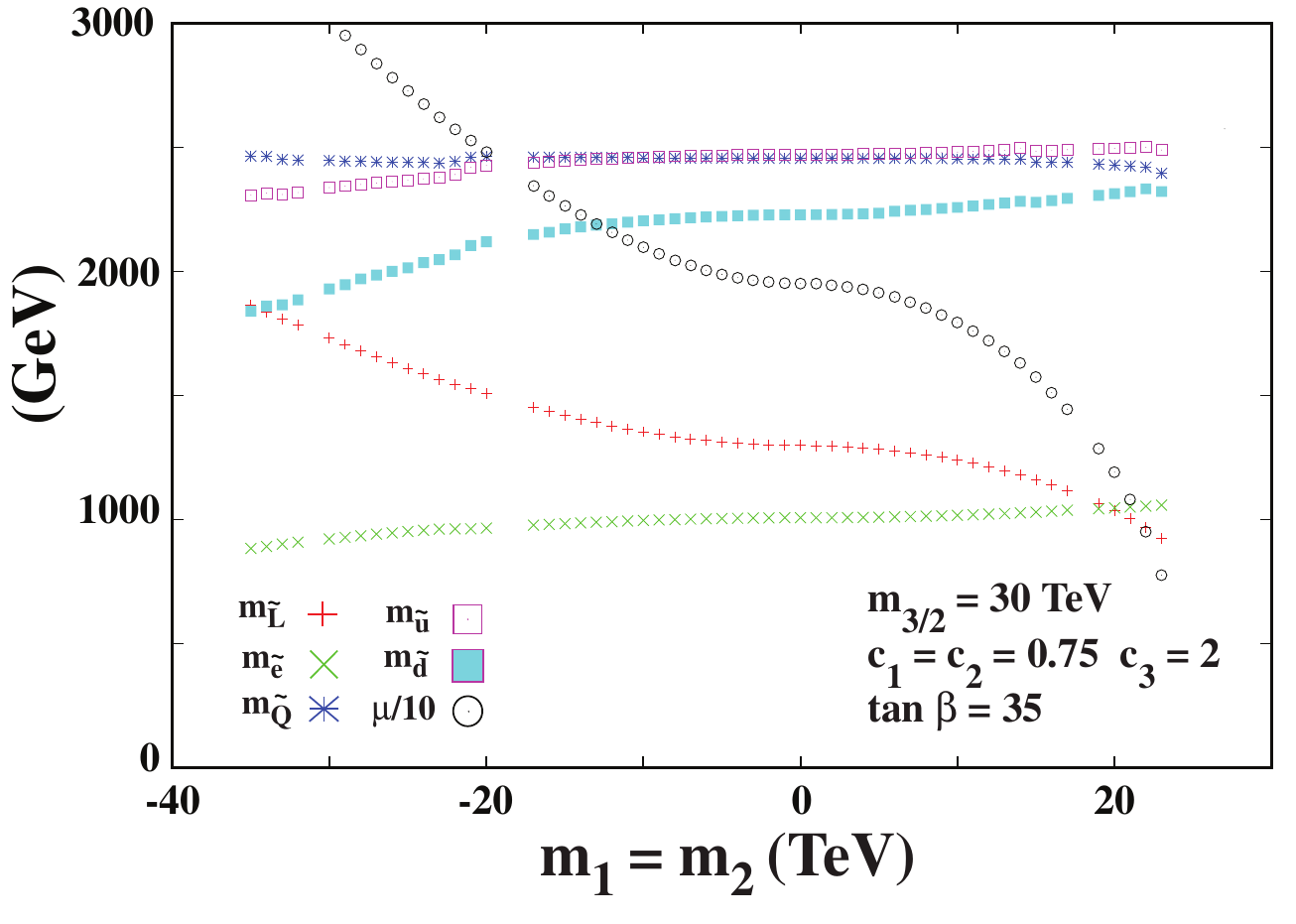,height=2.3in}
\end{minipage}
\begin{minipage}{8in}
\begin{center}
\hspace{-1.5in}\epsfig{file=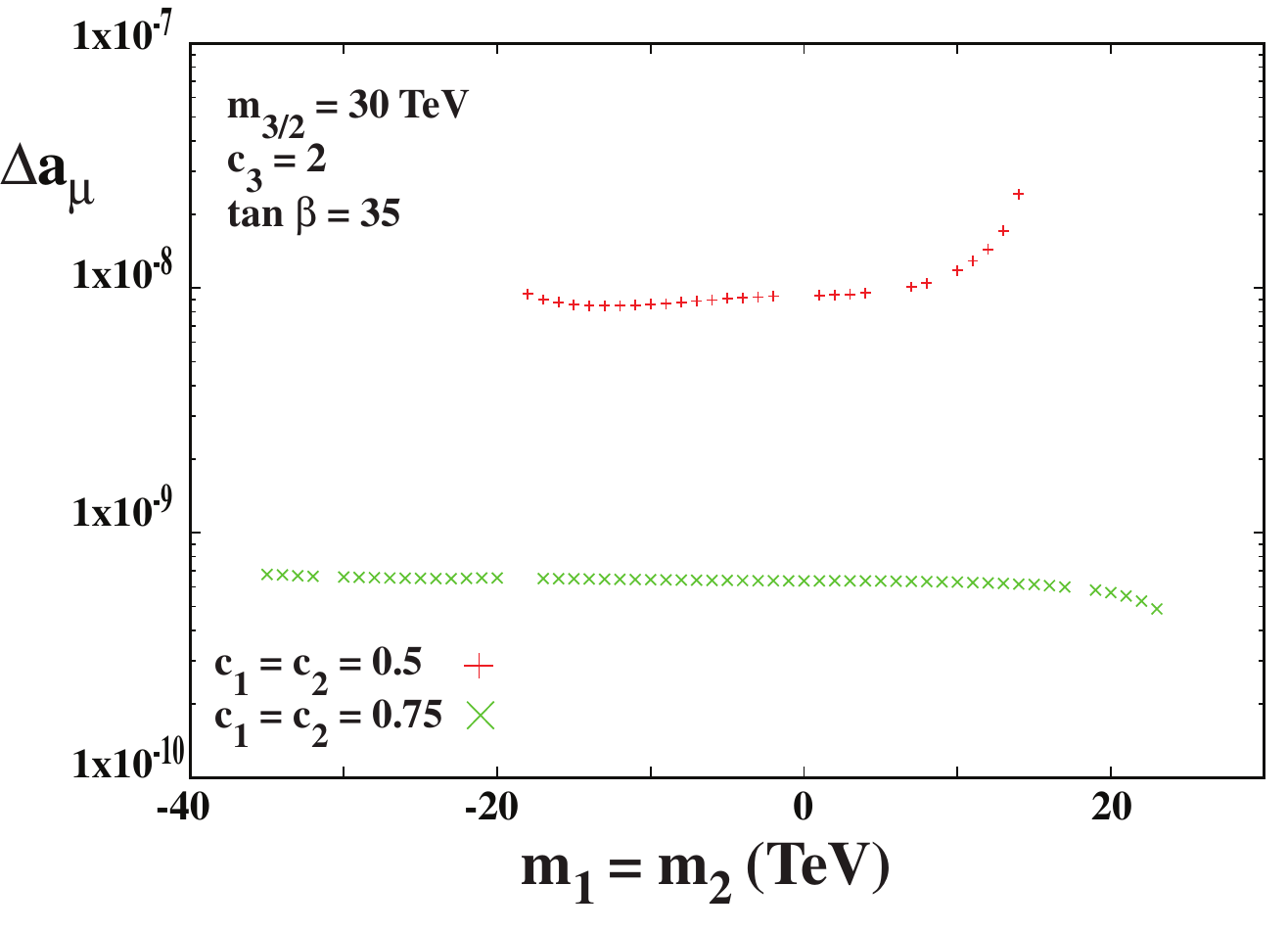,height=2.3in}
\end{center}
\end{minipage}
\caption{\it  \label{fig:m1m2vs}In  the top left and top right panels we show the mass spectra with respect to $m_1=m_2$ for $m_{3/2}=30$ TeV, $c_3=2$, $\tan\beta=35$, and $c_1=c_2=0.5$ and $c_1=c_2=0.75$ respectively. The red + is for the left-handed slepton. The green $\times$'s are for the right-handed sleptons. The blue star is for the left-handed squarks. The magenta box is for the right-handed up squark. The cyan filled box is for the right-handed down squark. The yellow circle is for $\mu/10$. In the lower panel, we have plotted $a_\mu$  for the same sets of parameters. The red +'s are for $c_1=c_2=0.5$ and the green $\times$'s are for $c_1=c_2=0.75$. }
\end{figure}

Because of the additional features of these models, a much lighter gravitino mass is allowed. In this case, the two-loop beta functions are much smaller and we can easily get a large enough correction to $(g-2)_\mu$ to explain the experimental discrepancy. Since $m_{3/2}$ is much smaller, we are free to take large $\tan\beta$. Here, we will take $c_1=c_2=1/2$, $c_3=2$, $m_{3/2}=30$ TeV, and $\tan\beta =35$ and scan over $m_1=m_2$.  We repeat this exercise for $c_1=c_2=3/4$.   The results of these scans can be seen in Fig. (\ref{fig:m1m2vs}).
As can be seen from the lower panel of Fig. (\ref{fig:m1m2vs}), the anomalous magnetic moment of the muon can be sufficiently enhanced with out tuning the $c_1,c_2$. Since the only parameters that are changing in these figures are the Higgs boundary masses, the Higgs mass is relatively unchanged and about $127$ GeV.

\begin{figure}
\begin{minipage}{8in}
\epsfig{file=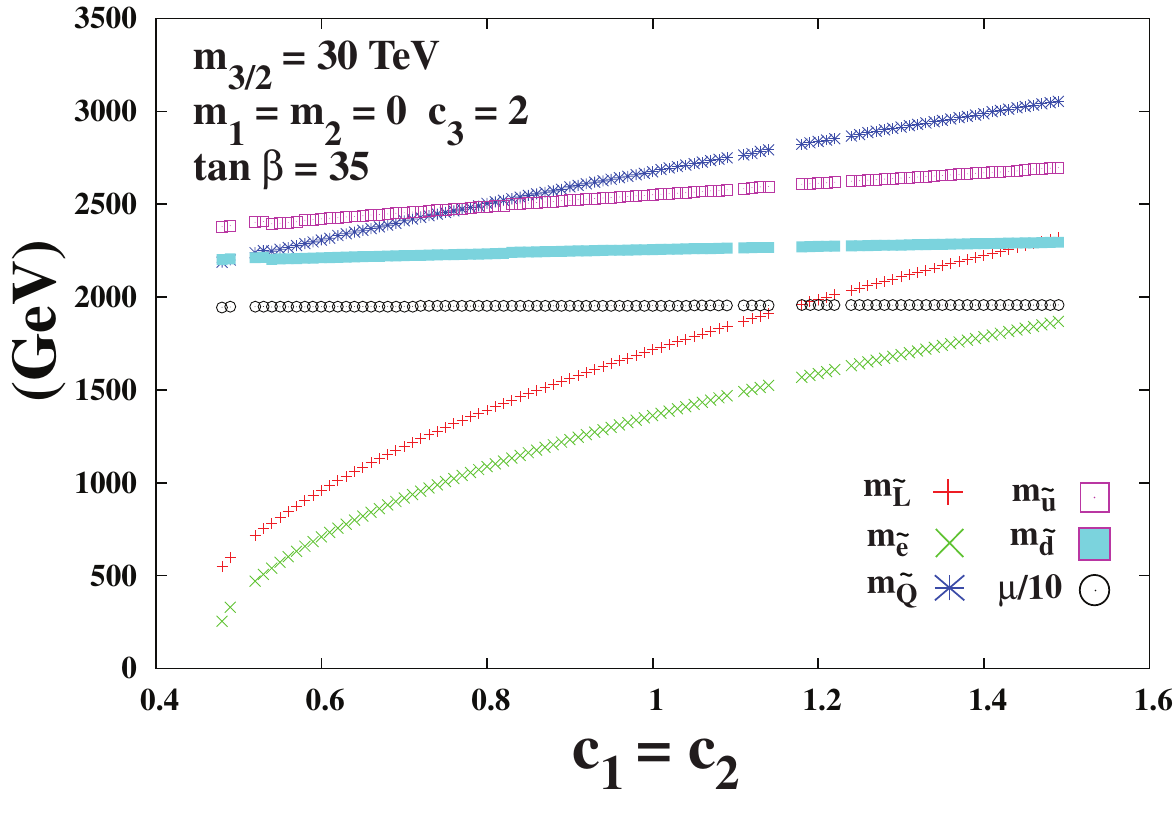,height=2.3in}
\epsfig{file=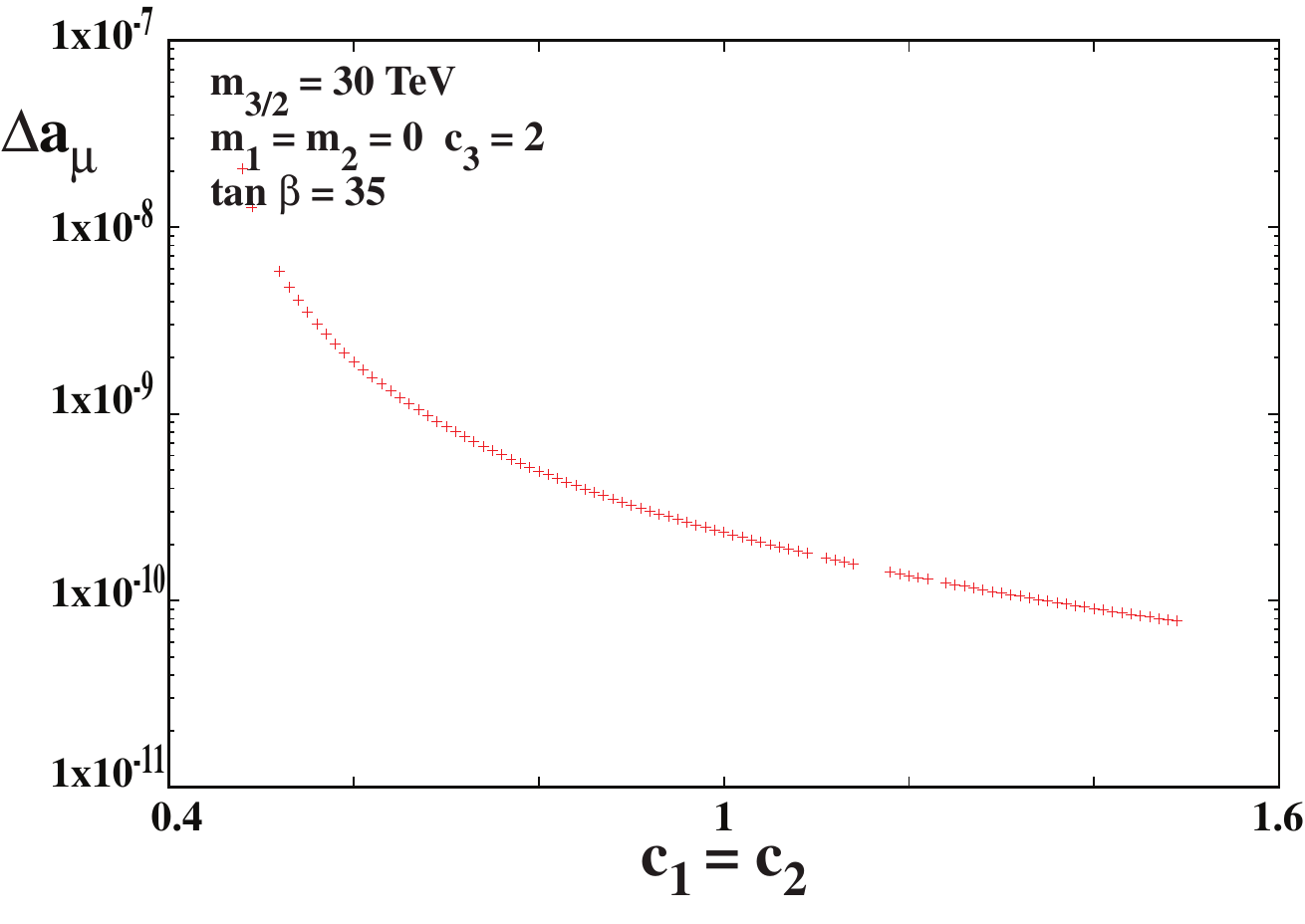,height=2.3in}
\end{minipage}
\caption{\it \label{fig:c1c2vs}In the left panel, we show the mass spectra for $m_{3/2}=30$ TeV, $c_3=2$, $\tan\beta=35$, and $m_1=m_2=0$. The symbols are as in Fig. (\ref{fig:m1m2vs}).  In the right panel, we have plotted $\Delta {\rm a}_\mu$  for the same set of parameters.  }
\end{figure}

Lastly, we plot the mass spectra and anomalous magnetic moment of the muon with respect to $c_1=c_2$, with $c_3=2$, $m_{3/2}=30$ TeV, $m_1=m_2=0$, and $\tan\beta=35$.  In Fig. (\ref{fig:c1c2vs}), we see that by varying $c_1,c_2$, we can easily get an anomalous magnetic moment consistent with experiment.

\section{Conclusions}
The recent discovery of the Higgs boson has placed rather severe constraints on simple models like the CMSSM.  To get a reasonable Higgs mass ($m_h > 124$ GeV) in the CMSSM, supersymmetry breaking mass parameters must be pushed to
order 1 TeV resulting in squark and gluino masses of order 2 TeV.
If however, there is a hierarchy between the sfermion masses and gaugino masses, such as in split-supersymmetry, PGM, and strong moduli stabilization, the observed Higgs mass can easily be accommodated. Furthermore, models such as PGM can be made consistent with radiative electroweak symmetry breaking for a limited range in $\tan \beta$. In addition, models with
strongly stabilized moduli tend to have a much simpler cosmology, avoiding the problems of excess entropy production and/or gravitino production \cite{ego}.  Indeed, for quite some time cosmological model building has suggested this hierarchy.

Although simple models like PGM have many advantages, there are some drawbacks to heavy sfermions.  If the squarks are heavy, detection at the LHC may be rather difficult. Furthermore, the deviation in $(g-2)_\mu$ has little hope of being explained, in this case.  Since both of these experimental difficulties hinge on the masses of the first two generations, while the Higgs mass depends primarily on the third generation masses, there may be hope of simultaneously getting all of these nice features. In fact, if the first and second generation masses are generated at one-loop, with respect to $m_{3/2}$ while the third generation masses remain at tree-level, both of these difficulties can be resolved.  In these scenarios, the down squark can be pushed below the gluino mass increasing the reach of the LHC for standard $SU(5)$ based models. These models also allow the sfermions to be light while keeping $\mu$ of order $m_{3/2}$. If the theory stems from product unification or has no unification at all, $(g-2)_\mu$ can be explained, even for sleptons of order $1$ TeV.

A nice and simple way to generate these one-loop masses is through anomaly like contributions.  If the regulated theory has Pauli-Villar fields which interact with the hidden sector, the theory will have one-loop masses generated by the gauge and Yukawa couplings \cite{oneloopmass}. The interactions of the Pauli-Villar fields with the hidden sector may be a natural part of string theory and by merely including this additional interaction at the Planck scale, we obtain one-loop masses.  Since we are quite ignorant about what the universe is like at the Planck scale, this is an acceptable assumption.

Lastly, we comment on the testable signatures of these models. One unique type of spectrum that can come from the type PGM we considered is a down squark that is lighter than the gluino. This unique spectra would result in an extended reach for the LHC and HLHC and would be fairly indicative of these types of models. If this form of PGM explains the deviation in the experimental value of $(g-2)_\mu$, we also expect that the wino should be seen at the ILC.  Otherwise, the sleptons would be too heavy to give a significant contribution to $(g-2)_\mu$. Although, these signatures are not necessary, they would be highly suggestive of this type of PGM model.

\appendix

\section{Off diagonal sfermion squared masses}
\label{sec:fcnc}
In the split family scenarios,  the model generically induces the FCNC processes through the flavor structure
of the Yukawa coupling (see e.g. Ref.\,\cite{Endo:2010fk}).
In our model, however, the FCNC contributions are  suppressed since
the soft masses in the first two generations are mainly generated by the one-loop anomaly mediated
contributions, and hence, are very close to each other.

In the soft mass diagonalized basis,  the mass terms and the supersymmetric Yukawa interaction terms
are  given by,
\begin{eqnarray}
{\cal L} &\simeq& m_{\tilde{f}0}^2 ( |\tilde{f}_1 |^2 + |\tilde{f}_2 |^2) + m_{\tilde f 3}^2  |\tilde{f}_3 |^2\ , \nonumber\\
W &=&  \bar{u}_R^i\, Y^u_{ij} Q_L^j H_u +  \bar{d}_R^i\, Y^d_{ij} Q_L^j H_u
+ \bar{e}_R^i\, Y^e_{ij} L_L^j H_u\ .
\end{eqnarray}
In this basis,
we expect
\begin{eqnarray}
|Y^{u,d,e}_{ij} | \ll | Y^{u,d,e}_{33} |\ , \quad ( i\neq 3\,\, {\mbox or}\, j\neq 3 ) \ .
\end{eqnarray}
when we assume that  the sfermion mass hierarchies are linked to the Yukawa coupling hierarchies.
It should be noted that the left-right mixing soft masses are safely neglected since they are suppressed
by the Higgs expectation value and by the small Yukawa couplings.

First, let us discuss the flavor mixing effects in the first two generations
at the tree-level.
For that purpose,  it is convenient to rotate the above scalar mass diagonal basis
into the so-called super-CKM basis by,
\begin{eqnarray}
 u_L^i = U_u^{i j} \tilde{u}_{Lj}\ ,  \quad
 d_L^i = U_d^{i j} \tilde{d}_{Lj}\ , \quad
 \bar{u}_R^i = V_u^{i j} \tilde{\bar u}_{Rj}\ ,  \quad
 \bar{d}_R^i = V_d^{i j} \tilde{\bar d}_{Rj}\ ,
\end{eqnarray}
where  the supersymmetric Yukawa couplings are diagonalized,
\begin{eqnarray}
 V_u^T Y^u U_u  = Y^u_{\rm diag} \ ,
 V_d^T Y^d U_d  = Y^d_{\rm diag} \ .
\end{eqnarray}
The CKM matrix is given by, $V_{\rm CKM} = U_u^\dagger U_d$.
In this basis, the soft squared masses have off diagonal elements,
\begin{eqnarray}
\label{eq:FCNC1}
m_{\tilde{f}ij}^2 = m_{\tilde f 0}^2 \delta_{ij} + X_{3i}^*X_{3j} (m_{\tilde f 3}^2 - m_{\tilde f0}^2 )\  ,
\end{eqnarray}
and similarly
\begin{eqnarray}
\left(m_{\tilde f ij}^{2}\right)^{-1} = \left(m_{\tilde f 0}^{2}\right)^{-1} \delta_{ij} + X_{3i}^*X_{3j} ((m_{\tilde f 3}^{2})^{-1} - (m_{\tilde f0}^{2})^{-1})\  ,
\end{eqnarray}
where $X = U_u, U_d, V_u^*, V_d^*$.
The mixing angles, $X_{31}$, $X_{32}$, are expected to be of ${\cal  O}(\lambda^3)$ and ${\cal O}( \lambda^2)$
with the Wolfenstein parameter $\lambda \simeq 0.2$, with respectively.
Therefore, from the above expression of $m_{ij}^{-2}$, we see that the flavor mixing parameter in the first two generations
is  of the order of  $|X_{31}X_{32}|$ at the tree-level, and hence, is highly suppressed.

Next, let us discuss the flavor violating effects from the RGEs.
In a general flavor basis, the flavor dependent part of the RGEs of the soft masses are given by%
\footnote{Flavor independent RGE contributions are absorbed in the $m_{\tilde{f}0}^2$ and $m_{\tilde f 3}^2$,
and hence, does not lead to an additional flavor mixing to the tree-level effects. }
\begin{eqnarray}
\frac{d}{dt} m_{Q\,ij}^2  &=&
  \frac{1}{16\pi^2}
    \left[ (m_{Q\,ik}^2 + 2 m_{H_u}^2 \delta_{ik}) Y^{u\dagger}_{k\ell} Y^u_{\ell j}
  +
    (m_{Q\,ik}^2 + 2 m_{H_d}^2 \delta_{ik}) Y^{d\dagger}_{k\ell} Y^d_{\ell j}\right.
\nonumber\\
&&      \left. +
 ( Y^{u\dagger}_{ik} Y^u_{k\ell } + Y^{d\dagger}_{ik} Y^d_{k\ell })    m_{Q\,\ell j}^2
 + 2 Y^{u\dagger}_{ik} m_{\bar u\,k \ell }^2  Y^u_{\ell j}
  + 2 Y^{d\dagger}_{ik} m_{\bar d\,k \ell }^2  Y^d_{\ell j} \right]\ ,
  \end{eqnarray}
  \begin{eqnarray}
\frac{d}{dt} m_{u\,ij}^2  =
  \frac{1}{16\pi^2}
    \left[ (2 m_{\bar u\,ik}^2 + 4 m_{H_u}^2 \delta_{ik}) Y^{u\dagger}_{k\ell} Y^u_{\ell j}
+
2  Y^{u\dagger}_{ik} Y^u_{k\ell }   m_{\bar u\,\ell j}^2
 + 4 Y^{u\dagger}_{ik} m_{Q\,k \ell }^2  Y^u_{\ell j}
    \right]\ ,
  \end{eqnarray}
\begin{eqnarray}
\frac{d}{dt} m_{\bar d\,ij}^2  =
  \frac{1}{16\pi^2}
    \left[ (2 m_{\bar d\,ik}^2 + 4 m_{H_d}^2 \delta_{ik}) Y^{d\dagger}_{k\ell} Y^d_{\ell j}
+
2  Y^{d\dagger}_{ik} Y^d_{k\ell }   m_{\bar d\,\ell j}^2
 + 4 Y^{d\dagger}_{ik} m_{Q\,k \ell }^2  Y^d_{\ell j}
    \right]\ .
  \end{eqnarray}
In the super-CKM basis, by neglecting the Yukawa couplings in the first two generations,
the above RGEs are reduced to
\begin{eqnarray}
\frac{d}{dt} m_{Q\,ij}^2  &\simeq&
  \frac{1}{16\pi^2}
  \left(
\begin{array}{ccc}
0  & 0  &y_t^2 m_{Q 13}^2  \\
0  & 0  &y_t^2 m_{Q23}^2   \\
y_t^2 m_{Q31}^2 &y_t^2 m_{Q32}^2  &
2y_t^2 (m_{H_u}^2 + m_{\tilde Q 3}^2+ m_{\tilde u 3}^2)
\end{array}
\right)\nonumber \ ,
 \end{eqnarray}
\begin{eqnarray}
\frac{d}{dt} m_{\bar u\,ij}^2  &\simeq&
  \frac{1}{16\pi^2}
  \left(
\begin{array}{ccc}
0  & 0  &2 y_t^2 m_{u13}^2  \\
0  & 0  &2y_t^2m_{u23}^2  \\
2y_t^2 m_{u31}^2 &2y_t^2m_{u32}^2 &
4y_t^2 (m_{H_u}^2 +  m_{\tilde Q 3}^2+  m_{\tilde u 3}^2)
\end{array}
\right)\nonumber \ ,
 \end{eqnarray}
\begin{eqnarray}
\frac{d}{dt} m_{\bar d\,ij}^2  &\simeq&
  \frac{1}{16\pi^2}
  \left(
\begin{array}{ccc}
0  & 0  &2 y_b^2 m_{d13}^2  \\
0  & 0  &2y_b^2 m_{d23}^2   \\
2 y_b^2 m_{d31}^2  &2 y_b^2 m_{d32}^2 &
4y_b^2 (m_{H_d}^2 +  m_{\tilde Q 3}^2+  m_{\tilde d 3}^2)
\end{array}
\right)\nonumber \ ,
 \end{eqnarray}
 Therefore, the soft  squared mass matrices in Eq.\,(\ref{eq:FCNC1}) receive flavor-violating radiative corrections.

By taking the inverse of the radiatively corrected soft mass squared at  the low energy scale, we  immediately find
 the radiatively induced flavor mixing parameters
\begin{eqnarray}
(\delta_{12}^d)_{LL} \sim \frac{y_{t}^2|X_{32} X_{31}|}{16\pi^2}\left(\frac{m_{\tilde f 3}^2}{m_{\tilde f 0}^2}\right) \log\frac{M_{\rm input}}{m_{\tilde f 3}} \simeq 10^{-3}
\left(\frac{X_{32} X_{31}}{10^{-5}}\right)
\left(\frac{m_{\tilde f 3}}{100\,\rm TeV}\right)^2
\left(\frac{3\,\rm TeV}{m_{\tilde f 0}}\right)^2\ ,
\end{eqnarray}
\begin{eqnarray}
(\delta_{12}^d)_{RR} \sim\frac{y_{b}^2|X_{32} X_{31}|}{16\pi^2}\left(\frac{m_{\tilde f 3}^2}{m_{\tilde f 0}^2}\right) \log\frac{M_{\rm input}}{m_{\tilde f 3}} \simeq 10^{-4}
\left(\frac{X_{32} X_{31}}{10^{-5}}\right)
\left(\frac{m_{\tilde f 3}}{100\,\rm TeV}\right)^2
\left(\frac{3\,\rm TeV}{m_{\tilde f 0}}\right)^2\ ,
\end{eqnarray}
at the leading order.
Here, we have estimated the radiative corrections in the leading log approximation.
We have also used  $y_b \simeq 0.2$  assuming $\tan\beta \simeq 10$.
As a result, the RGE induced FCNC contributions are also consistent with
the constraints from the $K_0 - \bar{K}_0$ mixing ($\Delta m_K $  and $\varepsilon$);
$((\delta_{12}^{d})_{LL}(\delta_{12}^{d})_{RR})^{1/2}\lesssim 10^{-3}(m_{\tilde d }/3\,{\rm TeV})$
and
$(\delta_{12}^{d})_{LL}\lesssim 10^{-2}(m_{\tilde d }/3\,{\rm TeV})$\,\cite{Gabbiani:1996hi}.

\section{Important Contributions to the Beta Function\label{sec:RGApp}}
The important contributions to the RG running of the sfermion masses come from loops of $D$ terms and so are proportional to gauge couplings.  The pure gauge contributions are\cite{Martin:1993zk}
\begin{eqnarray}
&&\Delta \beta_{m_{\tilde f}^2}^{g_Y}= \frac{Y^2}{3}\frac{g_Y^4}{(16\pi^2)^2}\left[ 3(m_2^2+m_1^2)+ {\bf Tr}\left( m_Q^2+3m_L^2 + 8m_u^2+2m_d^2+6m_e^2\right)\right]\label{eq:twoloopY}\\
&&\Delta \beta_{m_{\tilde f}^2}^{g_2}=\frac{3g_2^4}{(16\pi^2)^2}\left[ m_2^2+m_1^2+{\bf Tr}\left(3m_Q^2+m_L^2\right)\right]\label{eq:twoloopSU2}\\
&&\Delta \beta_{m_{\tilde f}^2}^{g_3}=\frac{16}{3}\frac{g_3^4}{(16\pi^2)^2}{\bf Tr}\left[2m_Q^2+m_u^2+m_d^2\right] \label{eq:twoloopSU3}\, .
\end{eqnarray}
The fourth and final important contribution comes from adding a loop to the one-loop $D$ term diagrams of hypercharge.  This contribution is \cite{Martin:1993zk}
\begin{eqnarray}
{\cal S}' =
\!\!\!\!\!\!\!\!&& \nonumber {\rm Tr} \left[
-(3 m_1^2 + m_Q^2) Y_u^\dagger Y_u + 4 Y_u^\dagger m_u^2 Y_u
+ (3m_2^2 - m_Q^2) Y_d^\dagger Y_d - 2 Y_d^\dagger m_d^2 Y_d+ (m_1^2 + m_{\ell}^2) Y_e^\dagger Y_e \right.\\
&& \!\!\!\!\!\!\!\!\left.  - 2 Y_e^\dagger m_e^2 Y_e \label{eq:twoloopSp}
\right]
 + \left [ \frac{3}{2} g_2^2 + \frac{3}{10}g_1^2 \right ]
\left\{ m_2^2 - m_1^2 - {\rm Tr} (m_{\ell}^2) \right\}
+ \left [ \frac{8}{3} g_3^2 + \frac{3}{2} g_2^2 + \frac{1}{30} g_1^2 \right ]
{\rm Tr} (m_Q^2 )\\
&&\!\!\!\!\!\!\!\!\nonumber -\left [ \frac{16}{3} g_3^2 + \frac{16}{15} g_1^2 \right ]
{\rm Tr} (m_u^2 )
+\left [ \frac{8}{3} g_3^2 + \frac{2}{15} g_1^2 \right ]
{\rm Tr} (m_d^2 )
+ \frac{6}{5} g_1^2 {\rm Tr} (m_e^2) \, ,
\end{eqnarray}
and contributes to the running of the sfermion masses as
\begin{eqnarray}
\Delta\beta_{m_{\tilde f}^2}^{{\cal S}'} = 2Y\frac{g_1^2}{(16\pi^2)^2}{\cal S}' \, ,
\end{eqnarray}
where $g_Y^2=(3/5) g_1^2$.

Here we give a naive estimate of the size of these contributions to the sfermion masses if we assume the third generation masses dominate.  We will also assume that the third generation masses do not run.  Although this is not true, it does give us a good order of magnitude estimate for the size of these contributions.  Since the Higgs soft mass running complicates our approximation, we will focus on the $SU(3)$ contribution.  In this approximation, we have
\begin{eqnarray}
\frac{64}{3}\frac{g_3^4}{(16\pi^2)^2}m_{3/2}^2 \, .
\end{eqnarray}
Using the RGE for the gauge couplings we find that this give
\begin{eqnarray}\Delta m_{\tilde f}^2 = -\frac{2}{9\pi^2} \left(g_3^2(\mu)-g_3^2(\mu_0)\right)m_{3/2}^2 \, ,
\end{eqnarray}
which is of order one-loop.  A similar but slightly less accurate calculation can be done for each gauge group with similar results.

\section{Toy Model with One-Loop Masses}
\label{sec:toy}
\begin{figure}[b!]
\begin{center}
\subfloat[PV Contribution]{
\includegraphics[width=.4\textwidth]{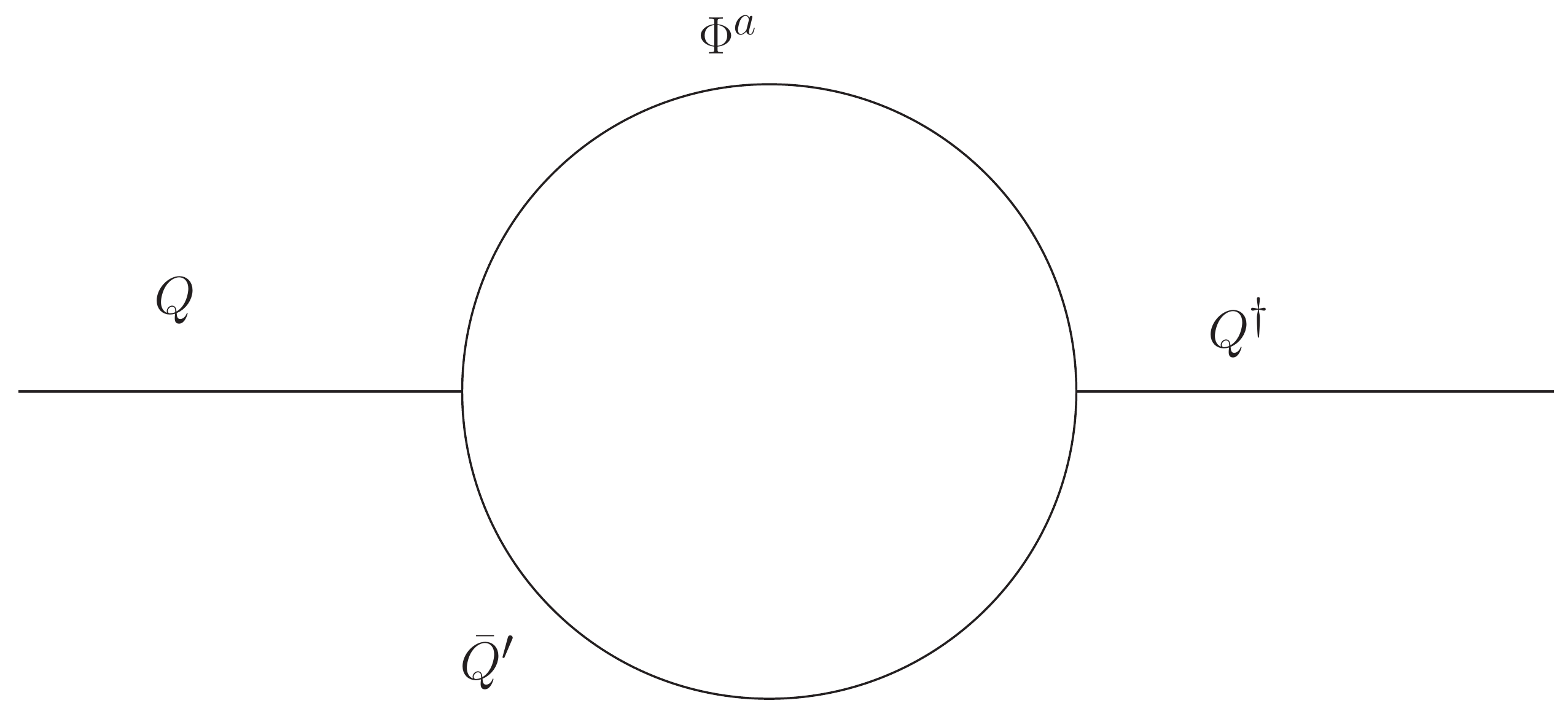}}
\subfloat[SUSY Contribution]{
\includegraphics[width=.4\textwidth]{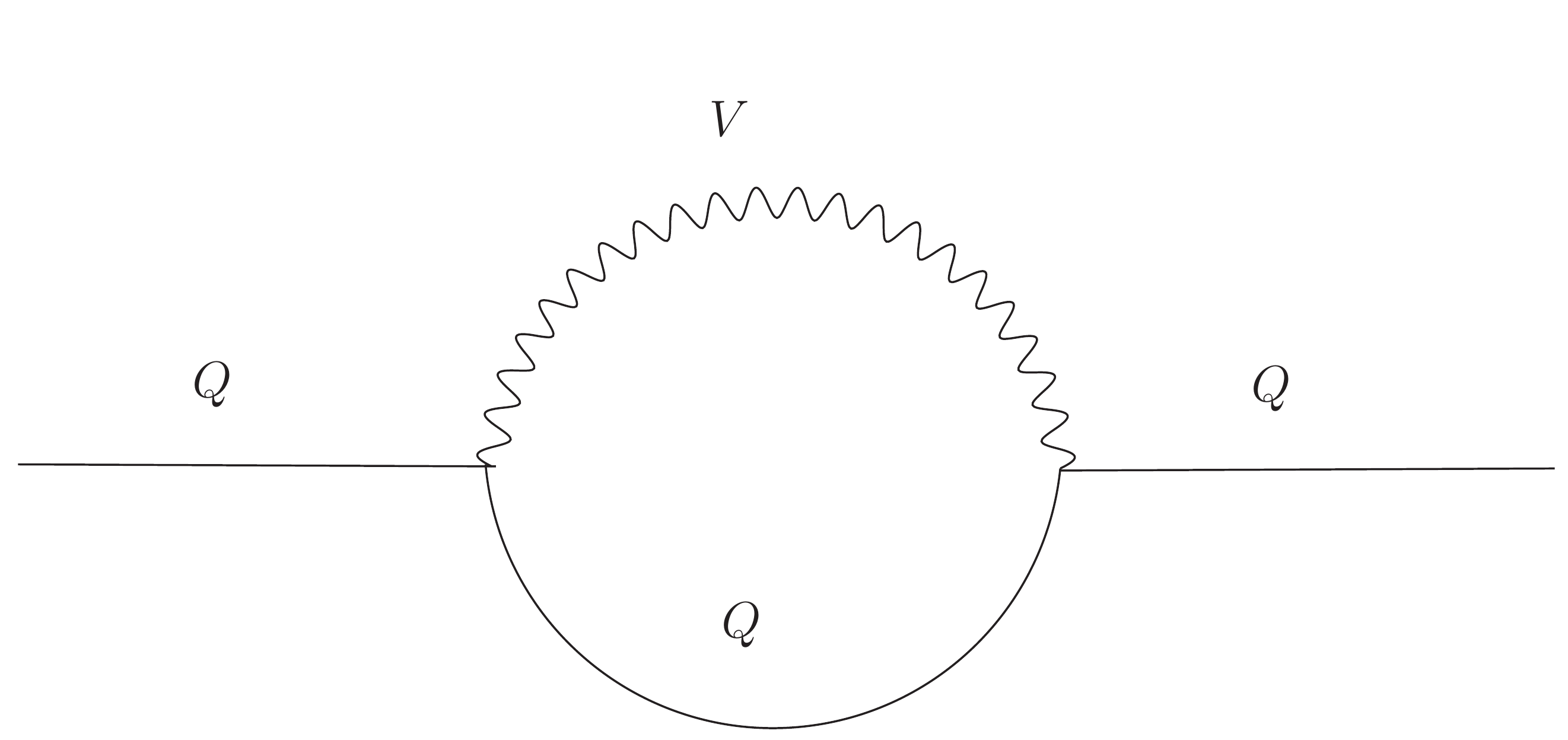}}
\caption{ The Feynman Diagrams for Renormalizing Yang-Mills theory at one-loop.  \label{1LRen}}
\end{center}
\end{figure}

Here, we provide a simple example of PV renormalization. This will be more of a sketch then a detailed calculation since we will not discuss the $\eta_i$ which will multiply each of the loops and are important for the cancellation of infinities.  These factors are needed to get the exact coefficients of the one-loop masses.  However, since this is not important to our considerations we will not address this issue and mostly focus on the diagrams themselves. We will also use a supergraph mass insertion method, which is perfectly justified since we assume the supersymmetric masses are much larger than the SUSY breaking masses.   We will start with the model
\begin{eqnarray}
K=Q^\dagger e^{V}Q \, ,
\end{eqnarray}
where the $Q$ are matter fields, and $V$  represents the gauge fields.
and we take no superpotential for the physical fields.  The K\"alher potential for the PV fields is
\begin{eqnarray}
K_{PV}=Q'^\dagger e^{V}Q'+\bar Q'^\dagger e^{V}\bar Q' + {\bf Tr}(\Phi^\dagger e^{V}\Phi) + {\bf Tr}(\bar \Phi^\dagger e^{V}\bar\Phi) \, ,
\end{eqnarray}
and superpotential
\begin{eqnarray}
W_{PV}=\mu\left( Q' \bar Q'+\Phi \bar \Phi\right) +g_i\sqrt{2} \bar Q' T^a \Phi^a Q \, ,
\end{eqnarray}
with $\Phi=\Phi^a T^a$ and the same for $\bar \Phi$. Now we calculate the one-loop wave function renormalization and the PV one-loop contribution that renormalizes it. The important graphs can be seen in Fig. (\ref{1LRen}) which is in supergraph notation. We have also suppressed the covariant derivatives.  To be clear, we will include an x for mass insertions in the graphs when we are referring to super propagators of the type $\langle \Phi \bar \Phi \rangle$ and  no x when we mean the propagators of the type $\langle \Phi^\dagger \Phi\rangle $.

\begin{figure}[b!]
\begin{center}
\subfloat[If $\bar Q'$ has a soft mass, likewise for $\Phi^a$]{
\includegraphics[width=.4\textwidth]{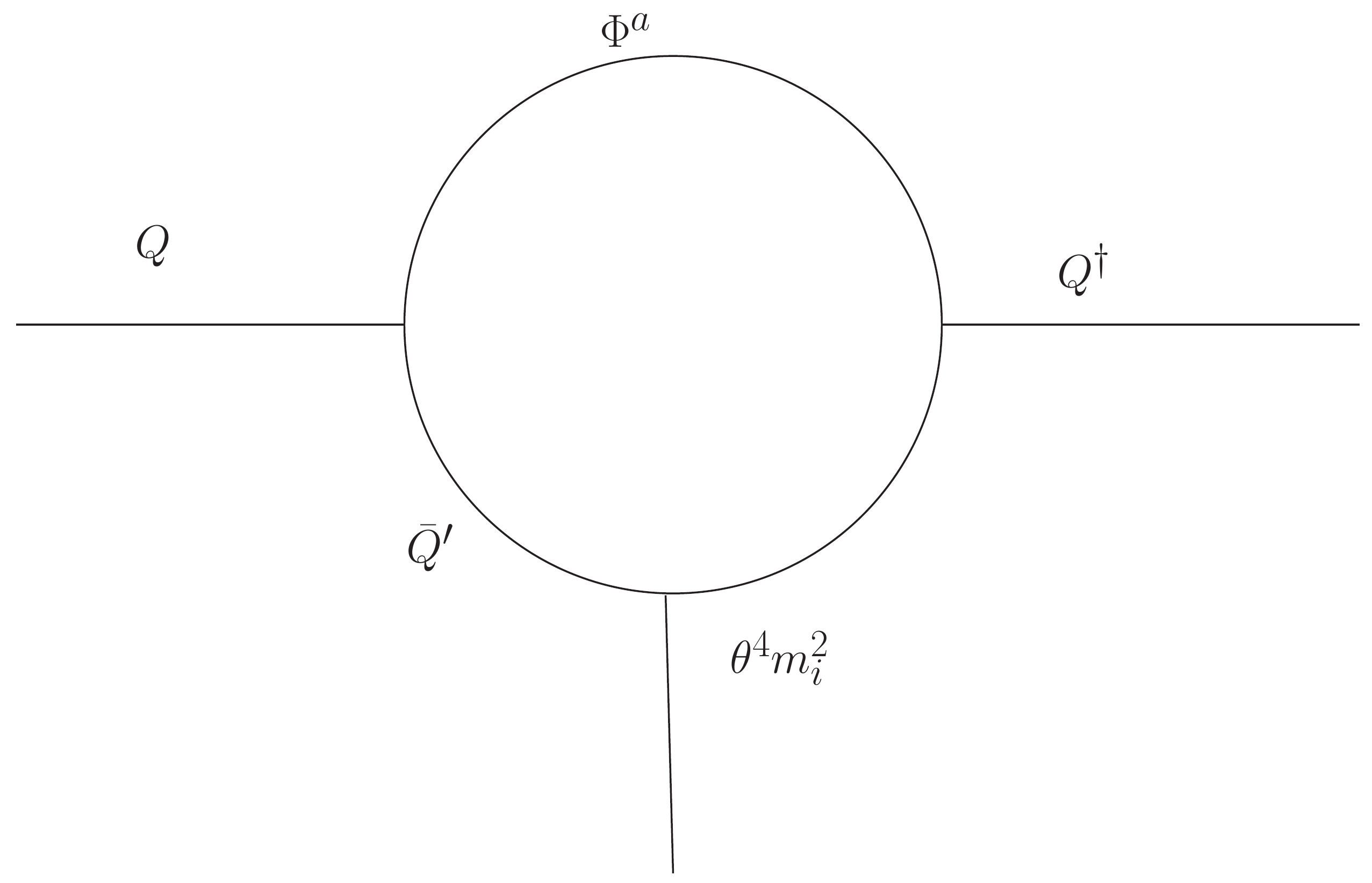}}
\subfloat[If $Q'$ has a soft mass, likewise for $\bar \Phi^a$]{
\includegraphics[width=.4\textwidth]{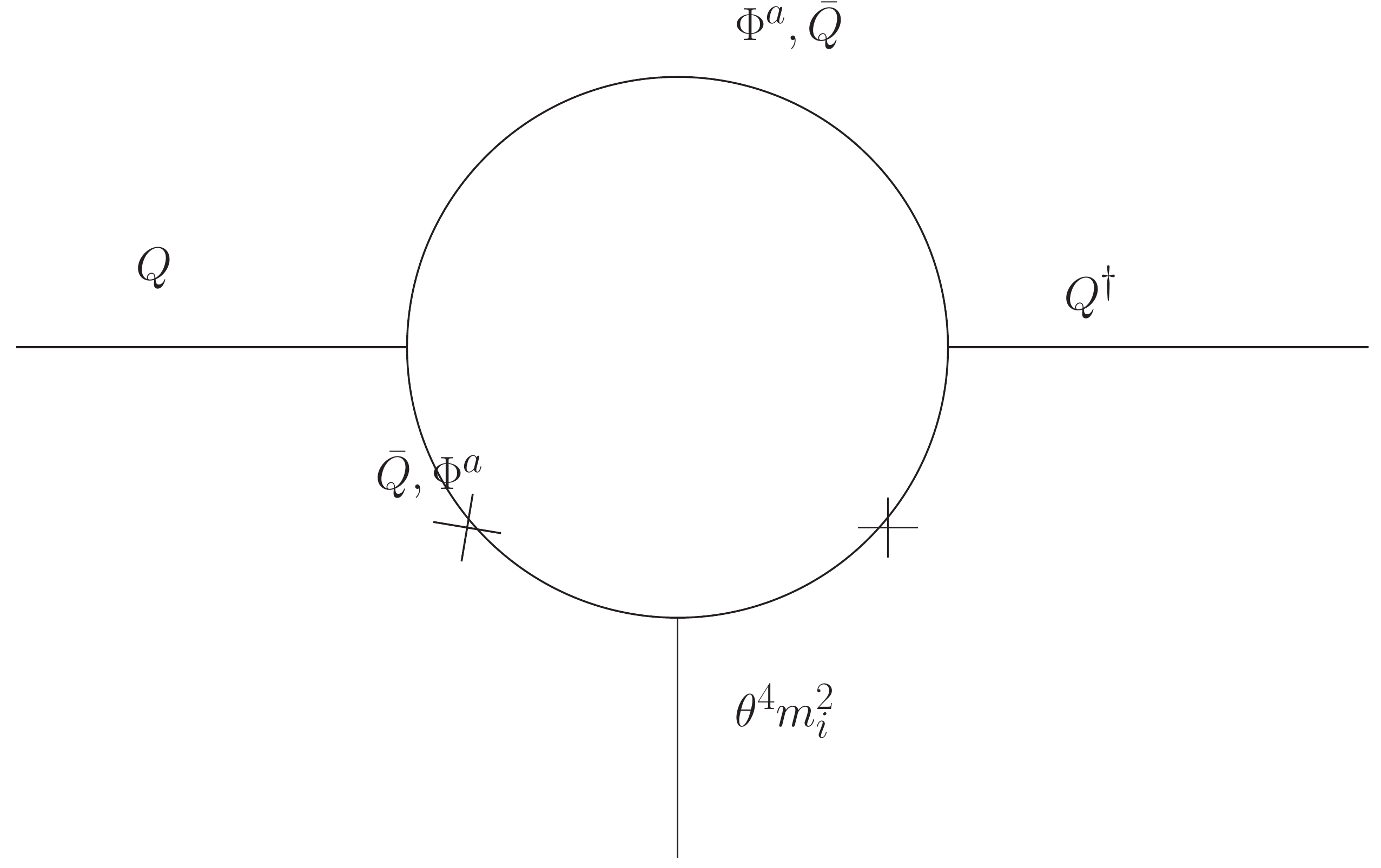}}
\caption{ The Feynman Diagrams which could give mass at one-loop. (left) This Diagram is only allowed if the physical field also has the same soft mass.  (right). This diagram gives a finite contribution and is always allowed.\label{SUBR}}
\end{center}
\end{figure}

If we calculate these graphs, we find a total contribution of
\begin{eqnarray}
\Delta K =2g_i^2C(r)\int d\theta^4\frac{d^4p}{(2\pi)^4} Q^\dagger (-p,\bar \theta)Q(p,\theta)\left( B_0\left(p^2,m_Q^2,0\right)-B_0\left(p^2,\mu,\mu\right)\right) \, ,
\end{eqnarray}
where
\begin{eqnarray}
B_0\left(p^2,m_1^2,m_2^2\right)= i\int \frac{d^4k}{(2\pi)^4}\frac{1}{k^2-m_1^2}\frac{1}{\left(p-k\right)^2-m_2^2} \,
\end{eqnarray}
Examining $\Delta K$, we see that this integral is indeed finite as long as $\mu$ is finite.  Since we don't care about the details of this renormalization we will just leave it at that.  Next we need to include SUSY breaking.  Looking at the graphs in Fig. (\ref{1LRen}), we can get some insight in to how the PV fields can act as messengers.  First, we comment on our calculation of the one-loop mass for $Q$.  In the example we are considering, $\bar Q '$ is one of the field running in the loop which renormalizes the gauge interactions. If $\bar Q'$ feels SUSY breaking differently than $Q$, the PV renormalization scheme would not work because there would be uncanceled infinities. The field $Q'$, on the other hand, never shows up in the one-loop supergraph.  Because of this fact, the theory can be renormalized nor matter how $Q'$ feels supersymmetry breaking. To show this, we will calculate two more graphs.  To incorporate SUSY breaking, we will modify the K\"ahler potential to read
\begin{eqnarray}
K_{PV} &=& (1+\theta^4 m_{\tilde Q'}^2)Q'^\dagger e^{V}Q'+(1+\theta^4 m_{\tilde{\bar Q}'}^2)\bar Q'^\dagger e^{V}\bar Q' \\
\nonumber  &+&(1+\theta^4 m_{\tilde \Phi '}^2) {\bf Tr}(\Phi^\dagger e^{V}\Phi) +(1+\theta^4 m_{\tilde{\bar \Phi}'}^2) {\bf Tr}(\bar \Phi^\dagger e^{V}\bar\Phi)
\end{eqnarray}
Using these corrections to the K\"ahler potential as interactions in our graphs we get the additional diagrams found in Fig. (\ref{SUBR}), which are again supergraphs.

With a mass insertion of $m_{\tilde{\bar Q}'}^2$, there is an additional contribution to one-loop renormalization of the $Q$ of
\begin{eqnarray}
\Delta K =2g_i^2C(r)m_{\tilde{\bar Q}'}^2\int d\theta^4\frac{d^4p}{(2\pi)^4} Q^\dagger (-p,\bar \theta)Q(p,\theta)\left( C_1\left(p^2,\mu^2,\mu^2,\mu^2\right)\right)
\end{eqnarray}
where
\begin{eqnarray}
C_1\left(p^2,m_1^2,m_2^2,m_3^2\right)= i\int \frac{d^4k}{(2\pi)^4}\frac{1}{k^2-m_1^2}\frac{k^2}{k^2-m_2^2}\frac{1}{\left(p-k\right)^2-m_3^2}
\end{eqnarray}
As can be seen from power counting, this is logarithmical divergent. Now, if $Q$ had a SUSY breaking mass of $m_{\tilde{\bar Q}'}^2$, we would get another diagram from the physical fields which would cancel this contribution. Clearly, we need these masses to be equal or our PV regularization doesn't work. Since we are considering the mass of the physical fields to be zero, this type of diagram will not appear. However, if we include a mass insertion of $m_{\tilde Q'}^2$ things are different.  Here, we need to change the propagators in the loop from $\langle \bar Q^\dagger Q\rangle $ to $\langle Q \bar Q\rangle \langle Q^{\prime \dagger} \bar Q^{\prime\dagger}\rangle m_{\tilde Q'}^2 \theta^4$. Doing this we get a mass contribution
\begin{eqnarray}
\Delta K =2g_i^2C(r)m_{\tilde Q'}^2\int d\theta^4\frac{d^4p}{(2\pi)^4} Q^\dagger (-p,\bar \theta)\theta^4Q(p,\theta)\left( C_0\left(p^2,\mu^2,\mu^2,\mu^2\right)\right)
\end{eqnarray}
where
\begin{eqnarray}
C_0\left(p^2,m_1^2,m_2^2,m_3^2\right)= i\int \frac{d^4k}{(2\pi)^4}\frac{1}{k^2-m_1^2}\frac{m_1m_2}{k^2-m_2^2}\frac{1}{\left(p-k\right)^2-m_3^2}
\end{eqnarray}
$C_1(p^2,\mu^2,\mu^2,\mu^2)$ is finite even in the limit $\mu\to \infty$ and can be easily evaluated giving
\begin{eqnarray}
\Delta K =2g_i^2C(r)\frac{m_{\tilde Q'}^2}{16\pi^2}\int d\theta^4\frac{d^4p}{(2\pi)^4} Q^\dagger (-p,\bar \theta)\theta^4Q(p,\theta) .
\end{eqnarray}
There is an identical diagram with a mass insertion in the $\Phi^a$ line.  Giving the same result with $m_{\tilde{\bar Q}'}^2 \to m_{\tilde \Phi}^2$. Summing these contributions to the mass of $Q$ we find
\begin{eqnarray}
m_{\tilde Q}^2 =g_i^2C(r)\frac{m_{\tilde{\bar Q}'}^2+m_{\tilde \Phi}^2}{8\pi^2}
\end{eqnarray}

This means we get a one-loop SUSY breaking soft mass no matter what the messenger scale is. Also, this is the exact one-loop contribution.  Any additional insertions of the soft mass in the diagram will lead to corrections of order $m_{\tilde f}^2/\mu^2$, which vanish when we take $\mu\to \infty$.  This is good since the PV fields mass should be taken to infinity in the end. Also, since the supersymmetric mass is much larger than the SUSY breaking masses for the PV fields, the sign of the soft mass does not need to be positive.  This means that the one-loop mass can be positive or negative.  This simple toy model gives results which are consistent with those found in \cite{oneloopmass}.

Although, we have applied this calculation to PV fields, it does have broader implications. For example, if the PV fields were take as just additional GUT scale fields, they would still generate one-loop masses if they interacted with hidden sector.  Therefore, we can generate one-loop masses from physical fields at any scale using this type of set up.

\section*{Acknowledgments}
We would like to thank N.Yokozaki for a useful discussion on the FCNC problem.
The work of J.E. and K.A.O. was supported in part
by DOE grant DE--FG02--94ER--40823 at the University of Minnesota.
This work is also supported by Grant-in-Aid for Scientific research from the
Ministry of Education, Science, Sports, and Culture (MEXT), Japan, No.\ 22244021 (T.T.Y.),
No.\ 24740151 (M.I.), and also by the World Premier International Research Center Initiative (WPI Initiative), MEXT, Japan.

\end{document}